\documentclass[multphys,vecphys]{svmult}

\usepackage{graphicx}    
\usepackage{multicol}    

\newcommand\beq{\begin{equation}}
\newcommand\eeq{\end{equation}}
\newcommand\bea{\begin{eqnarray}}
\newcommand\eea{\end{eqnarray}}

\begin{document}

\title*{Spin Wave Analysis of Heisenberg Magnets 
\\ in Restricted Geometries
\\ \small{[Lecture Notes in Physics v. 645, pp. 195-226 (2004)]}
}
\titlerunning{Spin Wave Analysis of Heisenberg Magnets 
\\ in Restricted Geometries}
\author{Nedko B.  Ivanov\inst{1}\and
Diptiman Sen\inst{2}}
\institute{Theoretische Physik II, Universit\"at Augsburg,
D-86135 Augsburg, Germany\footnote{Permanent address: Institute
of Solid State Physics, Bulgarian Academy of Sciences, 
Tsarigradsko chausse 72, 1784 Sofia, Bulgaria}
\texttt{ivanov@physik.uni-augsburg.de}
\and Center for Theoretical Studies, Indian Institute of Science,
Bangalore 560012, India \texttt{diptiman@cts.iisc.ernet.in}}
%
%
\maketitle

\abstract
In the last decade it has been proven  that the standard 
spin-wave  theory was able to provide accurate zero-temperature
results for a number of  low-dimensional Heisenberg spin systems. 
In this chapter we introduce  the main ingredients of 
the spin-wave technique  using  as a working model the two-leg 
mixed-spin ferrimagnetic ladder and the Dyson--Maleev boson formalism
up to second order in the spin-wave interaction.
In the remainder, we survey  typical applications 
in low-space dimensionality as well as some recent
modifications of the theory admitting  a quantitative analysis 
in magnetically  disordered phases.
The presented spin-wave  results are compared with available 
numerical estimates.

\setcounter{footnote}{0}
\section{Introduction}
\label{sec:1}
The spin-wave theory    
is probably one of the most powerful  tools 
ever used in the theory of magnetism. Originally proposed by Bloch 
\cite{bloch1,bloch2} and Holstein and Primakoff \cite{holstein}
as a theory of the  ferromagnetic state, 
it was later extended for the antiferromagnetic N\'eel state 
by Anderson \cite{anderson},  Kubo \cite{kubo}, and Oguchi \cite{oguchi}. 
Dyson's  profound  analysis of spin-wave interactions 
\cite{dyson1,dyson2} demonstrated that spin waves may be used to obtain
 asymptotic expansions for the thermodynamic functions of 
the Heisenberg ferromagnet at low temperatures.  
Dyson's method was  generalized by Harris et al. \cite{harris}
to calculate in a systematic way spin-spin correlations, 
spin-wave damping, and various thermodynamic properties 
of antiferromagnetic insulators. 

It should be noticed that
the basis of the spin-wave theory (SWT) for antiferromagnets 
is  much less established than for ferromagnets. 
The Dyson--Maleev transformation  \cite{maleev} gives a correspondence
between any operator defined on the Hilbert space of the spin system and 
an operator on the  boson Hilbert space. Evaluating the required averages
for the Bose system, we necessarily make two approximations.
First, we expand these quantities,  by using a perturbation formalism in 
which  the unperturbed Hamiltonian is quadratic in boson operators and
the perturbation is the remaining quartic interaction. Second, we
neglect the projection operator in the averages,  
which takes into account the so-called kinematic interactions by canceling 
the boson  states with more than $2S$ bosons per lattice site, $S$ 
being the spin quantum number of the lattice spin. In the ferromagnetic case,
Dyson has argued that  these  approximations would lead
to results which are asymptotically correct at low temperatures ($T$) 
to all orders in $T$. In the antiferromagnetic case, the situation 
is less settled  due to the zero-point motion, i.e. quantum spin
fluctuations in the  N\'eel state. In principle, one may suspect that 
there are  errors in the perturbation theory even at zero T. 
The same problem appears in the Holstein--Primakoff
 formalism  \cite{holstein}. We refer the interested 
reader to the   original 
papers cited above as well as to the monographs 
\cite{tyablikov,akhiezer,mattis} for details concerning this problem. 
In principle, the spin-wave approach is less effective  
for low-dimensional  quantum spin systems, 
as  quantum spin fluctuations  typically increase
in reduced  space  dimensions ($D$) and for  small spin quantum 
numbers  $S$.  Moreover,  since   at finite T
thermal fluctuations completely destroy the magnetic 
long-range order in  1D and 2D Heisenberg models 
with isotropic short-range interactions \cite{mermin}, 
in such cases the conventional SWT completely fails.

 In view of the mentioned drawbacks of SWT,   
it seems surprising  that  for the last decade 
the standard spin-wave   approach  has been found to give  very  
accurate description of the 
zero-temperature physics of a number of low-dimensional spin models, 
the best example  being the $S=\frac{1}{2}$ 
Heisenberg antiferromagnet on a square  lattice \cite{manousakis}.
Probably, another good example  is the 
mixed-spin Heisenberg chain
describing a large class of recently synthesized quasi-1D
molecular magnets \cite{ivanov1} (cf.  Chap. 4). 
The following analysis reveals some  common features of 
these examples, the most important being 
the weakness (in a sense) of  spin-wave
interactions. Fortunately, in low-space dimensions many 
numerical techniques --  such as the quantum 
Monte Carlo method (QMC), the exact numerical diagonalization (ED), and the
density-matrix renormalization group method (DMRG) -- 
are more effective, so that 
the discussed drawbacks of the spin-wave analysis may be partially 
reduced  by a direct combination with numerical methods.

A  goal  of the present review is to summarize   
typical applications and some recent developments  of the 
spin-wave approach  related to low-dimensional  
quantum spin systems. The spin-wave technique is presented 
in the following Sect.,  using   the mixed-spin Heisenberg ladder 
as a working model and the Dyson--Maleev boson formalism.
Due to the asymptotic character of  spin-wave series, 
the calculation  up to second  order in the spin-wave interaction  
is a  reasonable approximation  for most of the  applications 
at zero T. As far as  at this level perturbative 
corrections   can easily be calculated in the 
framework of the Rayleigh--Schr\"odinger theory, 
we will not consider  in detail perturbation techniques
based on magnon Green's functions \cite{harris,canali1}.
Typical   applications of the spin-wave 
formalism in low-dimensional  spin systems are presented
in Sects. 3 and 4. In particular, Sect.  3 involves 
an  analysis of  the   parameters of  the quantum 
ferrimagnetic phase in mixed-spin quasi-1D   models, 
such as  the $(s_1,s_2)$ Heisenberg chain. 
The SWT results are compared with available DMRG and 
ED numerical estimates. Section 4 collects  basic SWT results concerning
2D Heisenberg antiferromagnets. Some recent modifications of the SWT -- 
admitting  a quantitative analysis in magnetically  disordered phases --
are presented in Sect. 5.  Section  6  contains concluding
remarks.   
\section{Dyson--Maleev Formalism}
\label{sec:2}
\begin{figure}
\centering
\includegraphics[height=2.5cm]{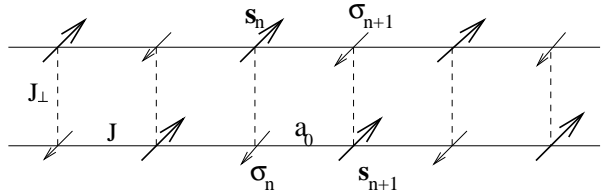}
%
%
\caption{Mixed-spin Heisenberg ladder composed of two types of site spins.
The arrows show one of the  classical ground states for   $J_{\perp}>0$,
defined  by the orientation 
of the ferromagnetic moment $\vec{M}=\sum_{n=1}^N
(\vec{s}_n+\vec{\sigma}_n)$.  The intrachain 
coupling $J=1$.}
\label{ladder}       
\end{figure}

In this Sect.  we describe the formal apparatus 
of the SWT. We choose as a working model the 
mixed-spin Heisenberg ladder
(Fig.~\ref{ladder}) defined by the Hamiltonian
\begin{equation}\label{h}
{\cal H}= \sum_{n =1}^{N}
\left[ \vec{s}_n\cdot \vec{\sigma}_{n+1}+\vec{\sigma}_n\cdot
\vec{s}_{n+1}\right]
+ J_{\perp}\sum_{n =1}^{N} \vec{s}_n\cdot \vec{\sigma}_n\, ,
\end{equation}
where the index $n$ ($=1,\cdots,N$) 
labels the rungs of the ladder, and $N$ is an even integer. 
The ladder is composed of two types
of spins ($\vec{s_n}$,$\vec{\sigma_n}$) 
characterized by the spin quantum  
numbers $s_1$ and $s_2$ ($s_1>s_2$):
$\vec{s_n}^2=\hbar^2s_1(s_1+1)$ and
$\vec{\sigma_n}^2=\hbar^2s_2(s_2+1)$. In the following text we 
use the notation $r_s\equiv s_1/s_2>1$, and set
$\hbar =1$ and $a_0 = 1$, $a_0$ being the lattice spacing
along the ladder.

It is worth noticing that the model (\ref{h}) is not 
purely  academic. For instance, recently 
published experimental
work on bimetallic quasi-$1D$ molecular magnets (cf. Chap. 4)
implies that the magnetic properties of these mixed-spin compounds
are basically described by the  Heisenberg spin model with
antiferromagnetically coupled nearest-neighbor localized spins.
The ladder structure in Fig.~\ref{ladder} reproduces,
in particular, arrangements of  the  Mn
($s_1=\frac{5}{2}$) and the Cu ($s_2=\frac{1}{2}$)
magnetic atoms along the $a$ axis in the compounds
MnCu(pbaOH)(H$_2$O)$_3$ (pbaOH = 2--hydroxy--1,3--propylenebisoxamato) 
\cite{kahn}.
\subsection{Classical Reference State}\label{sec:2.1}
The first step in  
constructing  a spin-wave expansion
is to find the lowest-energy classical spin configurations of the  related 
classical model. As a rule, this is a straightforward task,  
apart from some  magnetic models with competing interactions
which may exhibit complicated non-collinear  spin states 
(see, e.g.  \cite{diep}). Another serious problem at this stage 
may be related to a  macroscopic degeneracy  of the classical
ground state, a  typical example being the Heisenberg model on
a kagom\'e lattice (cf. Chap. 3) which exhibits a  magnetically 
disordered ground state. Further analysis of the problem involves
quantum fluctuations and  the so-called 
\emph{order-from-disorder phenomenon} \cite{shender,henley}.

Turning to  our model (\ref{h}), it is easy to see that
the required  reference state  
for  $J_{\perp}>0$ is  a ferrimagnetic
spin configuration where  the $\vec{s}_n$ spins are oriented 
in a given direction,  and the $\vec{\sigma}_n$ spins point in the
opposite direction (see Fig.~\ref{ladder}). The state is 
degenerate under arbitrary  rotations  (as a whole)  in the spin space. 
One may pick up a reference state   by  introducing  
a small  staggered field, say,   for  the  $\vec{s}_n$ spins. 
We can actually get  more information even in the quantum 
case, by  using the Lieb--Mattis theorem for bipartite lattices \cite{lieb}.
First, the theorem predicts  that the quantum ground state  belongs to
a subspace with the  total-spin quantum number
$(S_1-S_2)N$, i.e.  for $J_{\perp}>0$
the system has a ferrimagnetic ground state characterized by the
ferromagnetic moment per site $M_0=(s_1-s_2)/2$. 
Second, the theorem states   that  the energies of the
ground states $E(S_T)$ characterized by the total-spin 
quantum numbers  $S_T\ge N(s_1-s_2)$ are arranged as follows
\begin{equation}\label{es1}
E(S_T+1) > E(S_T)\, .
\end{equation}  
Notice that the classical and  quantum ferrimagnetic
ground states  have one and the same 
magnetization $M_0$, but otherwise they  are different because the 
classical ground state is not an eigenstate of the quantum model (\ref{h}).
The quantum  ferrimagnetic state is $[2N(s_1-s_2)+1]$-fold
degenerate, since  the $z$ component of the total spin -- being a
good quantum number -- takes the    
values $-N(s_1-s_2),-N(s_1-s_2)+1,\cdots,N(s_1-s_2)$.
This quantum magnetic phase may  also be characterized by the following 
sublattice magnetizations 
\begin{equation}\label{ma:mb}
\vec{m}_A=\frac{1}{N}\sum_{n=1}^N\langle \vec{s}_n\rangle
\hspace{0.3cm}
\vec{m}_B=\frac{1}{N} \sum_{n=1}^N\langle \vec{\sigma}_n\rangle\, ,
\end{equation}
where the symbol $\langle \cdots\rangle$ means a quantum-mechanical
average over the ground state. We shall later see  that
quantum spin fluctuations  
reduce the classical sublattice
magnetizations $s_1$ and $s_2$, but the magnetic long-range order
is preserved, i.e. $\vec{m}_A,\vec{m}_B\neq 0$. 

In the region $J_{\perp}<0$ the situation is different, i.e. 
the lowest-energy spin configuration is the  N\'eel  
antiferromagnetic state based on the composite rung spins $s_1+s_2$. 
Now the Lieb--Mattis theorem predicts that the exact quantum ground state
is a spin-singlet state, i.e. $S_T=0$ and $M_0=0$. Therefore,
it may be generally expected a magnetically disordered phase,
t.e.  $\vec{m}_A,\vec{m}_B= 0$, as  the isotropic Heisenberg
model (\ref{h}) is defined on a bipartite 1D  lattice
(see,  e.g. \cite{auerbach1}). In terms of the SWT this would mean that 
the classical antiferromagnetic state is swept out by  quantum
fluctuations, so that the concept of the spin-wave expansion does not
work at all. 
\subsection{Boson Hamiltonian}\label{sec:2.2}
Now we describe the second step in constructing
the spin-wave expansion, t.e. the transformation of (\ref{h}) to
a boson Hamiltonian. 
The most popular boson representation of
spin operators  has been suggested by Holstein and Primakoff \cite{holstein}.
Other useful representations have been  devised by
Schwinger  \cite{schwinger}, Maleev \cite{maleev}, 
Villain \cite{villain}, and Goldhirsch \cite{goldhirsch1,goldhirsch2}. 

We start by defining the Holstein--Primakoff representation 
for the spins $\vec{s}_n$ ($n=1,\ldots,N$):
\begin{eqnarray}\label{hpr}
s_n^+\! =\! \sqrt{2s_1}\,\sqrt{1-\frac{a_n^{\dag}a_n}{2s_1}}\, \,  a_n\, ,
\hspace{0.1cm}
s_n^-\! =\! \sqrt{2s_1}\,a_n^{\dag}\sqrt{1-\frac{a_n^{\dag}a_n}{2s_1}}\, ,
\hspace{0.1cm}
s_n^z\! =\! s_1\! -\! a_n^{\dag}a_n\, ,
\end{eqnarray}
where $s_n^{\pm}=s_n^x\pm s_n^y$ and  $s_1$ is the  spin quantum number.
$a_n$ and $a_n^{\dag}$ are  annihilation and creation 
boson operators satisfying the commutation relations
\begin{equation}\label{bcr}
[a_n,a_m^{\dag}]=\delta_{nm},\hspace{0.3cm}
[a_n,a_m]=[a_n^{\dag},a_m^{\dag}]=0\, .
\end{equation} 
Using the last equations, it is easy to show that the operators
defined by (\ref{hpr})  satisfy the commutation relations
for  spin operators 
\begin{equation}\label{spincr}
[s_n^+,s_n^-]=2s_n^z\, ,\hspace{0.5cm}
[s_n^z,s_n^{\pm}]=\pm s_n^{\pm}\, ,
\end{equation}
and the equation $\vec{s}_n^2=s_1(s_1+1)$.    
The operators  $a_n$ and $a_n^{\dag}$ act  in the 
infinite-dimensional boson Hilbert space spanned by the
orthonormal basis states     
\begin{equation}\label{basis}
|n_1,n_2,\ldots ,n_N) =\frac{
(a_1^{\dag})^{n_1}(a_2^{\dag})^{n_2}\cdots (a_N^{\dag})^{n_N}}
{\sqrt{n_1!n_2!\ldots n_N!}}|0)\, ,
\end{equation}
where $n_i$ ($=0,1,\ldots,\infty$) is the occupation number of site 
$i$. The reference vacuum state $|0)$ is defined 
by the relations $a_i|0)=0$ (for $\forall$ $i$).

It is possible to rationalize the square roots in (\ref{hpr}) by
the Maleev similarity transformation
\begin{equation}\label{maleev}
a_n\longmapsto \, \left( 1-\frac{a_n^{\dag}a_n}{2s_1}\right)^{1/2}a_n\, ,
\hspace{0.5cm}
a_n^{\dag}\longmapsto a_n^{\dag} 
\left( 1-\frac{a_n^{\dag}a_n}{2s_1} \right)^{-1/2} .
\end{equation}
This transformation is not unitary, but  preserves the number 
operator $a_n^{\dag}a_n$ as well as the commutation relations (\ref{bcr})
within the physically relevant Hilbert space ($n_i\leq 2s_1$ for 
$\forall$ $i$).
Applying the last transformation to (\ref{hpr}), we get 
the  
Dyson--Maleev boson representation
\begin{equation}\label{dma}
s_n^+=\sqrt{2s_1}\,  (1-a_n^{\dag}a_n/2s_1)\, a_n\, ,
\hspace{0.5cm}
s_n^-=\sqrt{2s_1}\, a_n^{\dag}\, ,
\hspace{0.5cm}
s_n^z=s_1-a_n^{\dag}a_n\, .
\end{equation}
Note that the operators $s_n^{\pm}$ in this representation are
not Hermitian conjugate in the boson space (\ref{basis}) so that 
in the general case they will generate non-Hermitian Hamiltonians.
Treatment of such Hamiltonians requires  some care, but
it seems that -- at least up to second order in the spin-wave 
interaction -- this does not cause serious problems.
More problematic is the
relation  between physical and unphysical states.
The latter  appear in the exact Holstein--Primakoff representation as
well, as  any actual calculation requires truncation 
of the asymptotic square-root series. 
Dyson's method \cite{dyson1}   eliminates
the  unphysical boson states by a projection operator
giving zero on these states. In practice, 
however, we are enforced to  eliminate this
operator. As already mentioned, this is the basic approximation 
of SWT. As a whole,  the Dyson--Maleev 
formalism has many advantages if one needs to go beyond 
the linear spin-wave  theory (LSWT) within a perturbation scheme. This is
because the  interactions between spin waves are better handled 
so that the unphysical singularities caused by the long-wavelength 
spin waves cancel out. 

To continue, we write a representation similar to (\ref{dma}) 
for  the spins $\vec{\sigma}_n$, by using  a new set of boson
fields  ($b_n$, $n=1,\ldots,N$):
\begin{equation}\label{dmb}
\sigma_n^+=\sqrt{2s_2}\, b_n^{\dag}\, (1-b_n^{\dag}b_n/2s_2)\, ,
\hspace{0.5cm}
\sigma_n^-=\sqrt{2s_2}\, b_n \, ,
\hspace{0.5cm}
\sigma_n^z=-s_2+b_n^{\dag}b_n\, .
\end{equation}
$b_n$ and $b_n^{\dag}$ satisfy the same commutation relations
(\ref{bcr}), and are supposed to commute with the set of $a$ bosons.
Here the reference state is chosen in the 
opposite direction, in accord with the classical spin configuration
in Fig.~\ref{ladder}.

Using  (\ref{dma}) and  (\ref{dmb}), we can find the 
boson image of any function of  spin 
operators. In particular, we are interested in the boson
representation of the spin Hamiltonian (\ref{h}), which we denote by
${\cal H}_B$. For the purposes of SWT, it is instructive to 
express ${\cal H}_B$ in terms of the Fourier transforms $a_k$ and
$b_k$ of the boson operators $a_n$ and $b_n$, by using the unitary
Fourier transformations
\begin{equation}\label{fourier}   
a_n=\frac{1}{\sqrt{N}}\sum_k {\E}^{\imag kn}a_k\, ,
\hspace{0.5cm}
b_n=\frac{1}{\sqrt{N}}\sum_k {\E}^{-\imag kn}b_k\, ,
\end{equation}
and the identity
$$
\frac{1}{N}\sum_{n=1}^N {\E}^{\imag (k-k')n}=\delta_{kk'}\, .
$$
It may be verified that this transformation is canonical, by showing
that the new operators $a_k$ and $b_k$ obey a set of commutation relations
identical to (\ref{bcr}). The wave vectors $k$ in the last expressions
are defined in the first Brillouin zone: 
$$
k=\frac{2\pi}{N}l\, ,\hspace{0.5cm}
l=-\frac{N}{2}+1,-\frac{N}{2}+2,\ldots,\frac{N}{2},\ .
$$
Notice that the rung spins  ($\vec{s}_n,\vec{\sigma}_n$) in 
Fig.~\ref{ladder} compose the $n$-th   magnetic 
(and lattice) elementary cell: 
this may be  easily observed by interchanging  the site spins of every (say) 
even rung in Fig.~\ref{ladder}.

We leave the Fourier transformation of ${\cal H}_B$ 
as  an exercise, and directly present the result in terms of
the new operators $a_k$ and $b_k$:
\begin{equation}\label{hb}
{\cal H}_B=-2\gamma_0r_sS^2+{\cal H}_0+V_{DM}^{'}\, ,
\end{equation}
where
\begin{equation}\label{h0}
{\cal H}_0= 2S
\sum_k\left[ \gamma_0\left(a_k^{\dag}a_k+r_s b_k^{\dag}b_k\right)
+\sqrt{r_s }
\gamma_k \left( a_k^{\dag}b_{k}^{\dag}+a_kb_{k}\right) \right]\, ,
\end{equation}
\begin{equation}\label{dm1}
V_{DM}^{'}\! =\! -\frac{1}{N}\sum_{1-4}\delta_{12}^{34}\!
\left( \! 2 \gamma_{1-4}a_3^{\dag}a_2b_1^{\dag}b_4\! +\!
\sqrt{r_s}\gamma_{1+2-4}a_3^{\dag}
b_2^{\dag}b_1^{\dag}b_4\! +\! \frac{1}{\sqrt{r_s}}\gamma_4a_3^
{\dag}a_2a_1b_4\! \right).
\end{equation}
Here $\gamma_k=J_{\perp}/2+\cos k$
($\gamma_0=J_{\perp}/2+1$), $\delta_{12}^{34} \equiv \Delta 
(k_1+k_2-k_3-k_4)$ is the \emph{Kronecker function}, and we have 
introduced the abbreviations  $(k_1,k_2,k_3,k_4)\equiv (1,2,3,4)$
for the wave vectors.

In a standard spin-wave expansion,  $1/s_1$ and  $1/s_2$
are  treated as small parameters, whereas the parameter  $r_s$
may be considered as a fixed number of order unity. In such a perturbation
scheme, it is convenient to set $1/S\equiv 1/s_2$ and use
$1/S$ as a small parameter. Thus,  the first term in (\ref{hb}) --
the classical ground-state  energy -- is proportional to $S^2$, 
the LSWT Hamiltonian ${\cal H}_0$  is multiplied by $S$, 
and the spin-wave interaction term $V_{DM}^{'}$ has the order ${\cal O}(1)$. 
We shall follow a  perturbation scheme  where
the diagonal terms of   $V_{DM}^{'}$, i.e. terms proportional to
the occupation-number operators $a_k^{\dag}a_k$ and  $b_k^{\dag}b_k$,
are treated together with ${\cal H}_0$  as a  zeroth-order  Hamiltonian,
whereas the rest of $V_{DM}^{'}$ is taken as a perturbation \cite{harris}.  
This is a more generic approach because  for some reasons
the spin-wave interactions may be weak  even
in the extreme quantum systems with $1/S=2$.  
\subsection{Quasiparticle Representation}\label{sec:2.3}

In the next step,  we diagonalize  the quadratic Hamiltonian
${\cal H}_0$, by using the Bogoliubov canonical transformation
 to
quasiparticle boson operators ($\alpha_k$ and $\beta_k$) \cite{holstein}:
\begin{equation} \label{bt}
a_k=u_k(\alpha_k-x_k\beta_k^{\dag})\, , 
\hspace{0.3cm}
b_k=u_k(\beta_k-x_k\alpha_k^{\dag})\, ,
\hspace{0.3cm}
u_k^2(1-x_k^2)=1\, .
\end{equation}

It is a  simple exercise to
find the transformation parameters $u_k$ and $x_k$ from the condition
which 
eliminates  the off-diagonal terms $\alpha_k\beta_k$ appearing in
${\cal H}_0$ after the transformation (\ref{bt}).
The result reads
\begin{equation} \label{uv}
u_k=\sqrt{\frac{1+\varepsilon_k}{2\varepsilon_k}}\, ,\hspace{0.3cm}
x_k= \frac{\eta_k}{1+\varepsilon_k}\, ,
\end{equation}
where
\begin{equation} \label{par} 
\varepsilon_k=\sqrt{1-\eta_k^2}\, ,
\hspace{0.3cm}
\eta_k=\frac{2\sqrt{r_s}}{r_s+1}
\frac{\gamma_k}{\gamma_0}\, .
\end{equation}
In  some  applications,  the quadratic Hamiltonian ${\cal H}_0$ 
may include  additional  ferromagnetic  bilinear terms 
(such as $a_k^{\dag}b_k$) so that the actual diagonalization 
is  more involved due to the increased number of  parameters (\ref{uv}).  
Some  diagonalization techniques for systems with  large  number 
of boson operators are presented  in \cite{tyablikov,colpa}.   

A quasiparticle representation of the quartic terms (\ref{dm1})
requires more technical work. As mentioned above, it is
instructive to pick up the quadratic diagonal terms  in $V_{DM}^{'}$
and to treat them together with  ${\cal H}_0$ as a zeroth-order 
approximation. A simple way to do  this is based on the
presentation  of $V_{DM}^{'}$ as a sum of normal-ordered  products  of 
boson quasiparticle operators. Apart from a constant,
the resulting expression for  $V_{DM}^{'}$
contains  diagonal and off-diagonal quadratic operator terms, and
normal-ordered quartic operator terms. We leave as an exercise  
this simple but somewhat cumbersome procedure and 
give  the final result for 
${\cal H}_B$ expressed in terms of the quasiparticle boson operators
 $\alpha_k$ and $\beta_k$:
\begin{equation}\label{hbb}
 {\cal H}_B =E_0+{\cal H}_D+\lambda V\, ,
 \hspace{0.5cm}   V=V_2+V_{DM}\, ,
 \hspace{0.5cm}\lambda \equiv 1\, .
\end{equation}

Here $E_0$ is the ground-state energy of the ferrimagnetic state
calculated up to the order ${\cal O}(1)$ in the 
standard $1/S$ expansion:
\begin{equation}\label{e0}
\frac{E_0}{N}=- 2\gamma_0r_sS^2-\gamma_0(1+r_s)\left(
1-\frac{1}{N}\sum_k\varepsilon_k\right)S
+{\rm e}_1+{\cal O}\left(\frac{1}{S}\right) \, ,
\end{equation}
where ${\rm e}_1=-2(c_1^2+c_2^2)-J_{\perp}(c_1^2+c_3^2)
-(2c_2+J_{\perp}c_3)c_1(r_s +1)r_s^{-1/2}$ and
\begin{equation}\label{cc}
c_1=-\frac{1}{2}+\frac{1}{2N}\sum_k\frac{1}
{\varepsilon_k}\, ,
\hspace{0.2cm}
c_2=-\frac{1}{2N}\sum_k\cos k
\frac{\eta_k}{\varepsilon_k}\, ,
\hspace{0.2cm}
c_3=-\frac{1}{2N}\sum_k\frac{\eta_k}
{\varepsilon_k}\, .
\end{equation}

${\cal H}_D$ is the  quadratic Hamiltonian 
resulting   from ${\cal H}_0$ and
the  diagonal  terms picked up from (\ref{dm1}):
\begin{equation}\label{hd}
{\cal H}_D=2S\sum_k\left[
\omega_k^{(\alpha)}\alpha_k^{\dag}\alpha_k
+\omega_k^{(\beta)}\beta_k^{\dag}\beta_k\right] \, ,
\end{equation}
where  up to ${\cal O}\left(1/S\right)$ the 
\emph{dressed dispersions} read
\begin{equation}\label{ok}
\omega_k^{(\alpha,\beta)}\!=\!
\gamma_0\left(\frac{r_s+1}{2}\varepsilon_k
\mp \frac{r_s-1}{2}\right)
\! +\! \frac{g_k^{\pm}}{2S}+{\cal O}\left(\frac{1}{S^2}\right)
\end{equation}
where $ g_k^{\pm}=(g_k\eta_k-d_0)\varepsilon_k^{-1/2}/2
\pm (r_s-1)(2c_2\! +\! c_3J_{\perp})r_s^{-1/2}/2$,
$g_k=2c_1(r_s+1)\gamma_kr_s^{-1/2}+
4c_2\cos k+2c_3J_{\perp}$, $d_0=4c_1\gamma_0+
(r_s+1)(2c_2+J_{\perp}c_3)r_s^{-1/2}$.

The functions $\omega_k^{(\alpha,\beta)}$ without ${\cal O}(1/S)$
corrections will be referred to as \emph{bare dispersions}.

Finally, the  quasiparticle interaction $V$
includes two different terms, i.e.  the  two-boson
interaction
\begin{equation}\label{v2}
V_2=\sum_k\left[V_k^{+}\alpha_k^{\dag}\beta_k^{\dag}
+V_k^{-}\alpha_k\beta_k \right]\, 
\end{equation}
defined by the vertex functions
\begin{equation}
V_k^{\pm}=\frac{d_0\eta_k-g_k}{2\varepsilon_k}
\mp\frac{r_s -1}{\sqrt{r_s}}c_1
\gamma_k\, ,
\end{equation}
and the quartic   
Dyson--Maleev interaction
\begin{eqnarray}\label{vdm}
V_{DM} &=&- \frac{J}{2N} \sum_{1-4} \delta_{12}^{34}
\Big[
V^{(1)}_{12;34} \alpha^{\dag}_1 \alpha^{\dag}_2 \alpha_3 \alpha_4 +
2V^{(2)}_{12;34} \alpha^{\dag}_1 \beta_2 \alpha_3 \alpha_4+
2V^{(3)}_{12;34} \alpha^{\dag}_1 \alpha^{\dag}_2 \beta^{\dag}_3 \alpha_4
\nonumber \\
&+& 4V^{(4)}_{12;34} \alpha^{\dag}_1 \alpha_3 \beta^{\dag}_4 \beta_2
+ 2V^{(5)}_{12;34} \beta^{\dag}_4 \alpha_3 \beta_2 \beta_1
+2V^{(6)}_{12;34} \beta^{\dag}_4 \beta^{\dag}_3 \alpha^{\dag}_2
\beta_1\nonumber \\
&+& V^{(7)}_{12;34} \alpha^{\dag}_1 \alpha^{\dag}_2 \beta^{\dag}_3
\beta^{\dag}_4+V^{(8)}_{12;34} \beta_1 \beta_2 \alpha_3 \alpha_4+
V^{(9)}_{12;34} \beta^{\dag}_4 \beta^{\dag}_3 \beta_2 \beta_1\Big]\, ,
\end{eqnarray}
defined by the vertex functions
$V^{(i)}_{12;34}$, $i=1,\ldots ,9$. We have adopted the
symmetric form of vertex functions used in \cite{canali1}.
The   explicit form of  $V^{(i)}_{12;34}$ depends 
on the concrete model. For  the ladder model (\ref{h}), the vertex functions
may be obtained from those of the Heisenberg
ferrimagnetic chain \cite{ivanov4}, using
the formal substitution  
$\cos k \longmapsto \cos k+J_{\perp}/2$.

In the following we shall  treat the spin-wave interaction
$V$  as a small perturbation to the diagonal Hamiltonian $E_0+H_D$.
To restore the standard $1/S$ series, one should (i) use
bare dispersion functions, and (ii) resume the series in
powers of $1/S$. 
\section{Spin Wave Analysis of  Quasi-1D Ferrimagnets}\label{sec:3}
In this Sect. we analyze 
the magnon spectrum 
and  basic   parameters of the  quantum 
ferrimagnetic phase of the model  (\ref{h}), by using the developed 
spin-wave formalism and  the  Rayleigh--Schr\"odinger  
perturbation theory up to second order in $\lambda$.
The SWT results are compared 
with available DMRG and ED numerical estimates.
\subsection{Linear Spin Wave Approximations}\label{sec:3.1}
In a standard 
linear spin-wave approximation we 
consider only the first two terms in (\ref{hb}), and discard 
 $V_{DM}^{'}$ as a next-order term in $1/S$. This corresponds to
the first two terms in the expression for the ground-state  
energy (\ref{e0}),
and to the first term in the expression for 
the quasiparticle dispersions 
(\ref{ok}). As a matter of fact, by using the normal-ordering
procedure, we have already got even the next-order
terms of the expansions in $1/S$ for these quantities.
\subsubsection{Magnon Excitation Spectrum}\label{sec:3.1.1}
\begin{figure}
\centering
\includegraphics[height=5.5cm]{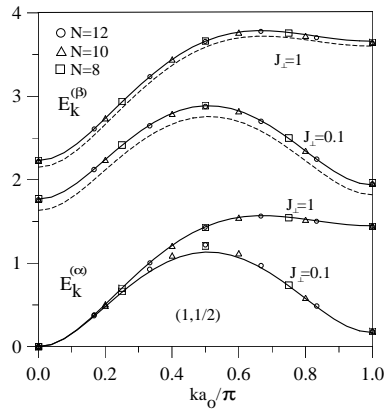}
\caption{Magnon excitation spectrum of the mixed-spin ladder $(s_1,s_2)=
(1,\frac{1}{2})$ for
interchain couplings $J_{\perp}=0.1$ and $J_{\perp}=1$. The dashed lines
display the   energy   of $\beta$  magnons $E_k^{(\beta)}$
related to the Hamiltonian ${\cal H}_D$. The solid lines
show  the magnon spectra as obtained from the  second-order  
approximation in  $V$. The  energy  of $\alpha$ magnons
related to (\ref{ok}) 
is not displayed, as it closely follows  the 
respective solid lines. The symbols indicate ED numerical results.
The Figure is taken from \cite{ivanov2}.
}
\label{spectrum}
\end{figure}

The quadratic Hamiltonian ${\cal H}_D$
defines two branches of
 spin-wave excitations
 ($\alpha$ and $\beta$ magnons)
described by the dispersion functions $\omega_k^{(\alpha,\beta)}$
in  the first Brillouin zone $-\pi \leq k \leq \pi$
(see Fig.~\ref{spectrum}).
The excited states  $\alpha_k^{\dag}|0\rangle$  
 ($\beta_k^{\dag}|0\rangle$) belong to the subspace characterized
by the quantum number $S_T^z= S_T-1$
($S_T^z=S_T+1$), where $S_T=(s_1-s_2)N$.
In the long wavelength limit $k\ll 1$, the energies of $\alpha$
magnons $E_k^{(\alpha)}$  have  the  
Landau--Lifshitz form
\begin{equation}\label{landau}
E_k^{(\alpha)}\equiv 2S\omega_k^{(\alpha)}=\frac{\varrho_s}{M_0}k^2
+{\cal O}(k^4)\, ,
\end{equation}
where  $\varrho_s$ is the spin stiffness constant \cite{halperin}.
This form of the Goldstone modes is typical for
Heisenberg ferromagnets,  and reflects the fact that the
order parameter, i.e.  the ferromagnetic moment,
is itself a constant of the motion.

 The spin stiffness constant $\varrho_s$ as well as $M_0$ 
play a basic role in the low-temperature thermodynamics \cite{read}.
The parameter  $\varrho_s$ may be obtained 
from the Landau--Lifshitz relation and (\ref{ok}):
\begin{equation}\label{rho0}
\frac{\varrho_s}{2s_1s_2}=1-\frac{1}{S}\left(
c_1\frac{r_s +1}{r_s}+\frac{c_2}{\sqrt{r_s}}\right)
+{\cal O}\left(\frac{1}{S^2}\right)\,  .
\end{equation}
The function $E_k^{(\alpha)}$    
exhibits an additional  minimum at the zone boundary,
so that in the vicinity of $\pi$ it reads
\begin{equation}\label{ll}
E_k^{(\alpha)}=\Delta_{\pi}^{(\alpha)}+{\rm const}
\left(\pi-k\right)^2\, .
\end{equation}
Here  $\Delta_{\pi}^{(\alpha)}$ is the excitation
gap at the zone boundary.
In the limit  $J_{\perp}\rightarrow 0$,
the excitation gap $\Delta_{\pi}^{(\alpha)}$ ($\propto J_{\perp}$) 
goes to zero. For ferromagnetic couplings $J_{\perp}<0$,
the $k=\pi$ mode  becomes unstable and produces 
global instability of the ferrimagnetic phase.

The function  $E_k^{(\beta)}\equiv 2S\omega_k^{(\beta)}$ may
be characterized by  the spectral gaps $\Delta_0^{(\beta)}$ (at $k=0$)
and $\Delta_{\pi}^{(\beta)}$ (at  $k=\pi$).
The expression for $\Delta_{0}^{(\beta)}$ reads
\begin{equation}\label{delta0}
\Delta_0^{(\beta)}=2\gamma_0(s_1-s_2)\left(1-
\frac{2c_2+c3J_{\perp}}
{2S\gamma_0\sqrt{r_s}}\right) 
+{\cal O}\left(\frac{1}{S}\right)\,  .
\end{equation}
For the $(s_1,s_2)=(1,\frac{1}{2})$ chain ($J_{\perp}=0$), 
the last equations give the results  $\varrho_s /2s_1s_2=0.761$ 
and $\Delta_0^{(\beta)}=1.676$, to be compared with 
the results $\varrho_s/2s_1s_2=1$ and $\Delta_0^{(\beta)}=1$ obtained 
in a standard linear approximation using
the Hamiltonian ${\cal H}_0$ \cite{pati1,brehmer}. 
A comparison  with the  numerical QMC
result  $\Delta_0^{(\beta)}=1.759$ \cite{yamamoto1}
clearly demonstrates the importance of the $1/S$ 
corrections to the dispersion functions (\ref{ok})
in the extreme quantum limit.

Summarizing, it may be stated that the linear
approximation -- based on the quadratic Hamiltonian ${\cal H}_D$ --  
gives  a good qualitative description of the magnon excitation 
spectrum of the  model (\ref{h}). The same conclusion is valid
for the ground-state  energy:  The expression  (\ref{e0})
has been found to produce an excellent fit to the 
numerical ED results in a large interval up 
to $J_{\perp}=10$ \cite{ivanov2}. 
\subsubsection{Sublattice Magnetizations}\label{sec:3.1.2}
The on-site magnetizations $m_A=\langle s_n^z\rangle$ 
and $m_B=-\langle \sigma_n^z\rangle$ are parameters of
the quantum ferrimagnetic phase which keep 
information for the long-range spin correlations. 
The simple LSWT results $m_A=s_1-c_1$ and $m_B=s_2-c_1$ 
show that quantum spin fluctuations  reduce the classical 
on-site magnetizations  already
at the level of non-interacting spin waves.  ${\cal H}_0$
produces the same results. The ratio 
\begin{equation}\label{measure}
\frac{s_2-m_B}{s_2}=\frac{c_1}{S}
\end{equation}
may be used as a measure of  the  
zero-point motion in the quantum ground state. Thus, there appears to be
a well-defined semiclassical limit $S\rightarrow \infty$ where
${\cal H}_0$   is a sufficiently 
accurate approximation, provided $c_1/S\ll 1$. 
In this connection, it seems surprising that  
the spin-wave series for the
$S=\frac{1}{2}$ square-lattice Heisenberg antiferromagnet
produces the excellent result $m_A=0.3069(2)$ \cite{hamer1} --
the recent stochastic-series QMC 
estimate is  $0.3070(3)$ \cite{sandvik} -- in spite of  the fact  that 
in this case  the parameter $c_1/S\approx 0.393$ is not small. 
Even more illuminating  is the $(1,\frac{1}{2})$ ferrimagnetic chain:
In spite of the large parameter $c_1/S\approx 0.610$, 
the second-order SWT gives the precise result 
$m_A=0.79388$ \cite{ivanov3} (DMRG estimate is  
$m_A=0.79248$ \cite{pati2}).
It is  difficult to explain the
accuracy of SWT in terms of the standard $1/S$ series.  
However, as will be shown below, 
the quasiparticle interaction $V$  produces  numerically 
small corrections to the principal zeroth-order approximation.
\begin{figure}
\centering
\includegraphics[height=5.5cm]{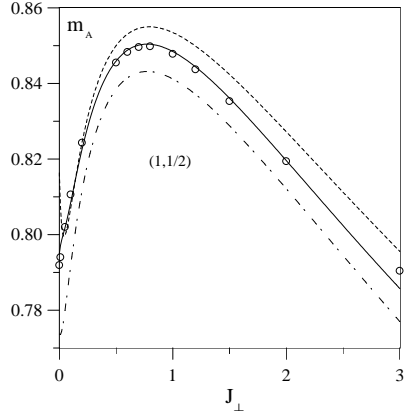}
\caption{
On-site magnetization (sublattice ${\cal A}$)
of the $(1,\frac{1}{2})$ ladder 
as a function of the interchain coupling $J_{\perp}$.
The dashed and dashed-dotted lines display the series results up to
first order in $1/S$ (bare dispersions) and $V$ (dressed dispersions).
The solid line shows the series result up to second order in $V$.
The Lanczos ED results for ladders with $N=12$ rungs are denoted by
open circles.
The Figure is taken from \cite{ivanov2}.
}
\label{slm}
\end{figure}

In  the mixed-spin model (\ref{h}) 
there appears an important first-order correction to the sublattice 
magnetizations which is  connected to the quadratic interaction $V_2$. 
Let us go beyond the linear approximation and calculate
the ${\cal O}(\lambda)$ correction  to $m_A$. The on-site
magnetization $m_B$ may  be 
obtained from the exact relation $m_A=s_1-s_2+m_B$ resulting from 
the conservation law for  the ferromagnetic moment. 
The expression of $m_A$ in terms of quasiparticle operators
reads
\begin{equation}\label{ma}
m_A=s_1-c_1-\frac{1}{2N}\sum_k\left[
\frac{1}{\varepsilon_k}\langle \alpha_k^{\dag}\alpha_k
+\beta_k^{\dag}\beta_k\rangle-
\frac{\eta_k}{\varepsilon_k}
\langle \alpha_k^{\dag}\alpha_k^{\dag}
+\beta_k^{\dag}\beta_k^{\dag}\rangle\right]\, .
\end{equation}
Now we  make use of the standard perturbation formula
\begin{equation}\label{form1}
\langle \hat{O} \rangle^{(1)}=\sum_{n\neq 0}\frac{\langle 0|
V|n\rangle\langle n|\hat{O}|0\rangle}{E_0-E_n}
+\sum_{n\neq 0}\frac{\langle n|
V|0\rangle\langle 0|\hat{O}|n\rangle}{E_0-E_n}\,  
\end{equation}
giving  the first-order correction in $V$ 
of $\langle \hat{O} \rangle$. Here
$\hat{O}$ is an arbitrary operator and
 $\langle \cdots \rangle$ means a quantum-mechanical 
average over the exact ground state.
 The formula is also valid in the case of 
non-Hermitian perturbations $V$.
In our case,  $\hat{O}$ is a quadratic  operator, so that  
the sum in (\ref{form1}) is restricted
to  the two-boson  eigenstates $|n_k\rangle = \alpha_k^{\dag}
\beta_k^{\dag}|0\rangle$ of ${\cal H}_D$, $k$
being a wave vector from the first Brillouin zone.
The  energies of these states are $E_k-E_0= 2S(\omega_k^{(\alpha)}
+\omega_k^{(\beta)})$. Finally, using  the  matrix elements
\begin{equation}\label{mel}
\langle 0|V_2|n_k\rangle =V_k^{(-)}\, ,\hspace{0.5cm} 
\langle n_k|V_2|0\rangle =V_k^{(+)}\, ,
\end{equation}
we get the following result
for $m_A$ calculated up to first order in $V$:
\begin{equation}\label{ma1}
m_A=s_1-c_1
-\frac{1}{4SN}\sum_k\frac{\eta_k}{\varepsilon_k}
\frac{V_k^{(+)}+V_k^{(-)}}{\omega_k^{(\alpha)}+\omega_k^{(\beta)}}
+{\cal O}(\lambda^2)\, .
\end{equation}
To find  the standard $1/S$ correction to $m_A$,
we have to use in (\ref{ma1}) the  bare  dispersion functions.

Figure~\ref{slm} shows  the results for $m_A$,  as obtained
from (\ref{ma1}) by using  the bare and  dressed dispersion 
functions (\ref{ok}). It is seen that the expansion in $1/S$ gives a
small (unexpected) decrease of $m_A$ in the vicinity of $J_{\perp}=0$,
whereas the expansion in $V$ produces a correct qualitative
result in this limit. The indicated problem of the standard $1/S$ series
probably results  from  enhanced fluctuations of the individual 
chain magnetizations about the common quantization axis. Indeed, at the 
special point $J_{\perp}=0$ the classical ground state 
acquires an additional degeneracy  under independent 
rotations of the chain ferromagnetic  
moments. Thus, the quartic diagonal interaction -- included in 
${\cal H}_D$ -- seems to  stabilize the common quantization axis
connected to the global ferromagnetic moment. 
We have an example where the expansion in powers of 
$V$ gives better results. 
\subsubsection{Antiferromagnetic Chain}\label{sec:3.1.3}
It is instructive to consider  the  antiferromagnetic chain as  a 
special case ($s_1=s_2,J_{\perp}=0$) of the 
mixed-spin model (\ref{h}).  After some algebra, from
(\ref{e0}) and (\ref{ok}) we find the following simplified 
expressions  for  the ground-state energy (per site)
\begin{equation}\label{af:en}
e_0=-S^2\left[1+\frac{1}{2S}\left( 1-\frac{2}{\pi}\right)\right]^2
+{\cal O}\left(\frac{1}{S}\right)\, 
\end{equation}
and    the magnon spectrum  
\begin{equation}\label{af:sp}
 \omega_k^{(\alpha,\beta)}\equiv \frac{E_k}{2S}
=\left[ 1+\frac{1}{2S}\left(
1-\frac{2}{\pi}\right)\right]|\sin k|+
{\cal O}\left(\frac{1}{S^2}\right)\, 
\end{equation}
of  the  antiferromagnetic chain.
For $S=\frac{1}{2}$,  the standard  LSWT  gives
the result  $e_0=-0.4317$ which is close to
Hulthen's exact result $-\ln 2+1/4\approx -0.443147$ \cite{hulthen}. 
It is an illuminating agreement, as the theory might 
have been expected  to fail for magnetically disordered states. 
Notice, however, that  the next-order approximation, 
i.e. $e_0=-0.4647$, does not improve the SWT result. 
This indicates a poor convergence  of the $1/S$ expansion. 
We can also check the series for $S=\frac{3}{2}$, by using 
the numerical result $e_0=-2.82833(1)$ \cite{hallberg} 
based on DMRG estimates for finite systems and the finite-size 
corrections to the energy,
as derived from the Wess-Zumino-Witten theory \cite{affleck1}.   
The first two terms in the series  (\ref{af:en}) for $S=\frac{3}{2}$   give
the result  $e_0=-2.79507$. In this case, an  inclusion of 
the  next-order term in  (\ref{af:en})
produces the precise  SWT result  $e_0=-2.82808$.
Thus, already for $S=\frac{3}{2}$ the spin-wave series shows 
a good convergence.

Turning to the magnon spectrum (\ref{af:sp}), we find that
for $S=\frac{1}{2}$ SWT  qualitatively reproduces 
Des Cloizeaux and Pearson's exact result for the one-magnon triplet
excitation spectrum $E_k=\frac{\pi}{2}|\sin k|$ \cite{cloizeaux}.
It is interesting  that the $1/S$ correction in (\ref{af:sp})
improves the  standard LSWT result  for the spin-wave velocity ($c=1$)
to  the value $c=1.3634$:  the exact result   
is $c=\pi/2\approx 1.5708$.
The magnon spectrum (\ref{af:sp}) is doubly degenerate
and has the relativistic form $E_k=c|k|$ ($c|\pi-k|$) near the 
point $k=0$ ($k=\pi$), to be compared with the rigorous result
where the spin-wave states, being eigenstates of spin 1,
are triply degenerate. Long-wavelength spin waves correspond to
states where all regions are locally in a N\'eel ground state
but the direction of the sublattice magnetization makes long-wavelength 
rotations. 

Using (\ref{cc}) and (\ref{measure}), we find the following 
expression for the  on-site magnetization in the 
antiferromagnetic chain  
\begin{equation}\label{measure1}
m=S-c_1=S+\frac{1}{2}-\frac{1}{2N}\sum_k\frac{1}{|\sin k|}=-\infty\, .
\end{equation}
We see  that in $1D$ the quantum correction  is divergent at small 
wave vectors already in the  leading LSWT approximation, no matter 
how large is $S$. This indicates  that the N\'eel state is
destabilized by quantum fluctuations, so that the concept of
spin-wave expansion fails.  

Finally, it is instructive to calculate the long-wavelength behavior of
the correlation function $\langle \vec{s}_n\cdot \vec{\sigma}_{n+x}\rangle$. 
Using  the Dyson--Maleev  representation
 and (\ref{bt}), one  finds 
$\langle \vec{s}_n\cdot \vec{\sigma}_{n+x}\rangle =
-S^2+2S\langle a_nb_{n+x}\rangle +\cdots$  where 
$\langle a_nb_{n+x}\rangle =-(1/2N)\sum_k (\cos k/|sin k|)\exp (\imag kx)$.
Thus,  in the limit $x\gg 1$ one obtains
\begin{equation}
\langle \vec{s}_n\cdot \vec{\sigma}_{n+x}\rangle
= -S^2\left[1-\frac{1}{\pi S}\ln x+{\cal O}
\left( \frac{1}{S^2}\right)\right].
\end{equation}
This indicates that in the semiclassical limit 
$S\rightarrow \infty$ the antiferromagnetic chain  is ordered 
at exponentially large scales $\xi \simeq a_0\exp (\pi S)$
\cite{affleck2}. Here we have restored the lattice spacing $a_0$.
\subsection{Spin Wave Interactions}\label{sec:3.2}
We have already discussed  some effects  of the 
quasiparticle interaction $V$, by calculating the first-order
correction to the sublattice magnetizations $m_A$ and $m_B$.
Notice that ${\cal O}(\lambda)$  corrections to the ground-state
energy (\ref{e0}) as  well as to the dispersion functions (\ref{ok}) 
do not appear. Indeed, it is easy to see  that the corresponding 
matrix elements $\langle 0|V|0\rangle$ and $\langle n_k|V|n_k\rangle$
($|n_k\rangle =\alpha_k^{\dag}|0\rangle$, or $\beta_k^{\dag}|0\rangle$)
vanish as a result of  the normal ordering  of $V$. 
It will be shown below  that the ${\cal O}(\lambda^2)$
corrections lead to  further improvement of the spin-wave results. 
To that end, we consider  two examples, i.e.  
the ground-state energy $E_0$ and the dispersion function 
$\omega_k^{(\alpha)}$. The reader is referred
 to the original literature  for similar
calculations concerning the parameters $m_A$, $\varrho_s$ \cite{ivanov2}, 
and $\Delta_0^{(\beta)}$ \cite{ivanov4}. 

The calculations may be performed 
within  the standard perturbation formula
\begin{equation}\label{form2}
E_i^{(2)}=\sum_{j\neq i}\frac{\langle i|V|j\rangle
\langle j|V|i\rangle}{E_i-E_j}
\end{equation}
giving the second-order correction in $V$ to the eigenvalue
$E_i$ of the eigenstate  $|i\rangle$ of a non-perturbed Hamiltonian. 
In our case, the   zeroth-order  Hamiltonian
is  $E_0+{\cal H}_D$, and the perturbation $V$ is given by (\ref{hb}). 
The sum in (\ref{form2}) runs over the eigenstates of ${\cal H}_D$. 
\subsubsection{Second-Order Corrections to $E_0$}\label{sec:3.2.1}
We consider corrections to the vacuum state $|i\rangle \equiv |0\rangle$,
so that the energy  $E_i\equiv E_0$ is given by (\ref{e0}). 
There are two types of ${\cal O}(\lambda^2)$ corrections to $E_0$
which are connected with the interactions 
$V_2$ and $V_{DM}$.

First, we  proceed with  the quadratic interaction $V_2$. 
It is easily seen that  only the   states $|j\rangle\equiv
|n_k\rangle = \alpha_k^{\dag}
\beta_k^{\dag}|0\rangle$  produce
 non-zero matrix elements in (\ref{form2}).
The dominator for these two-boson
states reads $E_0-E_k=-2S(\omega_k^{(\alpha)}+\omega_k^{(\beta)})$,
where  $\omega_k^{(\alpha,\beta)}$ are  defined by (\ref{ok}). 
Using the  above results and (\ref{mel}), we get the following 
correction to the ground-state energy 
(\ref{e0}) coming from $V_2$:
\begin{equation}\label{e21}
{E_0^{(2)}}^{'}=-\frac{1}{2S}\sum_k
\frac{V_k^{(+)}V_k^{(-)}}{\omega_k^{(\alpha)}+\omega_k^{(\beta)}}\, .
\end{equation}

Next, we  consider the Dyson--Maleev interaction $V_{DM}$.
Looking at the explicit expression of $V_{DM}$ (\ref{vdm}),
we find that only  the term with the vertex function 
$V^{(7)}_{12;34}$ ($V^{(8)}_{12;34}$) does not annihilate 
the vacuum state $|0\rangle$ ($\langle 0|$).
Thus, the sum in (\ref{form2}) runs over the four-boson eigenstates
$|1234\rangle =(2!2!)^{-1/2}
\alpha_{k_1}^{\dag}\alpha_{k_2}^{\dag} 
\beta_{k_3}^{\dag}\beta_{k_4}^{\dag}|0\rangle$. 
The related  matrix elements read
$$
\langle 1234|V_{DM}|0\rangle =-\frac{1}{N}V^{(7)}_{12;34}
\delta_{12}^{34}\, ,\hspace{0.5cm}
\langle 0|V_{DM}|1234\rangle =-\frac{1}{N}V^{(8)}_{43;12}
\delta_{12}^{34}\, .
$$
Using these expressions,  we find  the following 
correction to the ground-state energy resulting from $V_{DM}$:
\begin{equation}\label{e22}
{E_0^{(2)}}^{''}=-\frac{1}{2S}\frac{1}{N^2}\sum_{1-4}\delta_{12}^{34}
\frac{V^{(8)}_{43;12}V^{(7)}_{12;34}}
{\omega_1^{(\alpha)}+\omega_2^{(\alpha)}
+\omega_3^{(\beta)}+\omega_4^{(\beta)}}\, .
\end{equation}
Notice  that  the second-order correction to $E_0$ in powers of
$1/S$ is the sum of ${E_0^{(2)}}^{'}$ and ${E_0^{(2)}}^{''}$ but
 calculated with  the  bare dispersion functions.
\subsubsection{Second-Order Corrections to 
$\omega_k^{(\alpha)}$}\label{sec:3.2.2.}
\begin{figure}
\centering
\includegraphics[height=5cm]{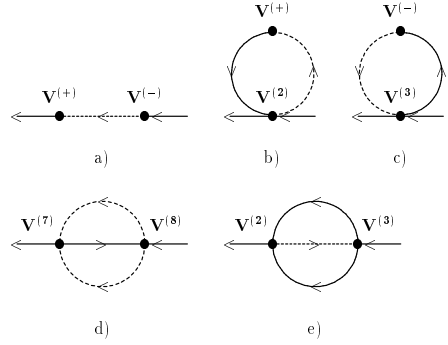}
\caption{
Second-order self-energy diagrams giving the corrections
to the dispersion function $\omega_k^{(\alpha)}$.
Solid and dashed lines represent, respectively, the bare 
propagators for $\alpha$ and  $\beta$ magnons.
The Figure is taken from \cite{ivanov4}.
}
\label{diagram}
\end{figure}

Now  we are interested in perturbations  to the one-magnon states
$|i\rangle \equiv |k\rangle =\alpha_k^{\dag}|0\rangle$.  
The calculations  may be performed by
following the method already used for $E_0$. 
Since we are  treating  an excited  eigenstate, 
there  appear new types of corrections  connected to  
the vertex functions   $V^{(2)}_{12;34}$ and
$V^{(3)}_{12;34}$. These terms  may be predicted, e.g. 
by drawing  the diagrams shown in Fig.~\ref{diagram}. 
Notice that the graphical  representation of the vertex functions in 
Fig.~\ref{diagram} is connected to the quasiparticle 
operator forms  of $V_2$ (\ref{v2}) and $V_{DM}$
(\ref{vdm}). The interested reader is referred to the original
literature (see, e.g. \cite{harris,canali1,baym}) where this 
diagram technique is explained in detail.      
We leave these  simple  calculations 
as  an exercise, and directly  present the expression
for the  second-order corrections to $\omega_k^{(\alpha)}$:
\begin{eqnarray}\label{corr:oma}
\!\!\!&&\delta \omega^{(\alpha)}_k= -\frac{1}{(2S)^2}
\left[ \frac{V^{(+)}_{k}V^{(-)}_{k}}
{\omega^{(\alpha )}_k+\omega^{(\beta )}_k }
-\frac{2}{N}\sum_{p}\frac{V^{(+)}_{p}V^{(2)}_{kp;pk}
+V^{(-)}_{p}V^{(3)}_{kp;pk}}
{\omega^{(\alpha )}_p+\omega^{(\beta )}_p }\right.\nonumber \\
\!\!\!&&+\frac{2}{N^2}\left.
\! \! \sum_{2-4}\delta_{k2}^{34}
\! \left(  \frac{V^{(8)}_{43;2k}V^{(7)}_{k2;34}}
{\omega^{(\alpha )}_k+\omega^{(\alpha )}_2+\omega^{(\beta )}_3+
\omega^{(\beta )}_4 }\!
+\! \frac{V^{(3)}_{43;2k}V^{(2)}_{k2;34}}
{-\omega^{(\alpha )}_k+\omega^{(\beta )}_2+\omega^{(\alpha )}_3+
\omega^{(\alpha )}_4 }\right)\! \right] .\hspace{0.6cm}
\end{eqnarray}

It is interesting to note   that
the vertex functions $V_k^{(-)}$, $V^{(2)}_{kp;pk}$,
$V^{(3)}_{kp;pk}$, $V^{(8)}_{43;2k}$, and $V^{(3)}_{43;2k}$
vanish at the zone center $k=0$ \footnote{Analytical properties 
of the vertex functions
have been studied in \cite{kopietz}}, so that
the gapless structure of $\omega_k^{(\alpha)}$ is
preserved separately by each of the second-order corrections
in (\ref{corr:oma}). Thus, we have  an  example  demonstrating
some of the good features of  the Dyson--Maleev  formalism.
\subsection{Comparison with  Numerical  Results}\label{sec:3.3}
We have already presented in Figs.~2 and 3
second-order SWT results for the dispersion functions 
$\omega_k^{(\alpha,\beta)}$
and the on-site magnetization $m_A$ of the $(1,\frac{1}{2})$
ladder. The comparison shows that the SWT dispersion 
functions closely follow   
the ED data in the whole Brillouin zone. For instance, 
the SWT result for the gap $\Delta_0^{(\beta)}$ at $J_{\perp}=0.1$
differs by less than $0.5\%$ from the ED estimate. Turning to
$m_A$, we find  a precision higher than  $0.3\%$  
in the whole interval $0\leq J_{\perp}\leq 3$. These are illuminating
results, as in the considered system the perturbation parameter 
$1/S=2$ is large.  
To  understand these results, 
let us consider, e.g.    the $\lambda$
series for the spectral gap $\Delta_0^{(\beta)}$ of the
 $(1,\frac{1}{2})$ chain \cite{ivanov4}:
$$
\frac{\Delta_0^{(\beta)}}{2(s_1-s_2)}=1.6756\lambda^0+0.1095\lambda^2
-0.0107\lambda^3+{\cal O}(\lambda^4)\, .
$$
Although $1/S=2$, we  see that the quasiparticle 
interaction $V$ introduces numerically small corrections to 
the zeroth-order approximation  ${\cal H}_D$.
\begin{table} 
\centering
\caption{Spin-wave results for the  parameters ${\rm e}_0=E_0/N$,
$m_A$, and $\Delta_0= \Delta_0^{(\beta)}/2(s_1-s_2)$ of 
different $(s_1,s_2)$  Heisenberg chains  calculated, respectively,
 up to the orders $1/S$, $1/S^2$, and $1/S^3$.  The SWT results are
compared with available DMRG estimates   which 
are, respectively,  denoted by  $\bar{\rm e}_0$, $\bar{m}_A
$\cite{pati2}, and  
$\bar{\Delta}_0$ \cite{ono}.
}
\begin{tabular}{lllllll}
\hline\noalign{\smallskip}
($s_1,s_2$) &\hspace{0.2cm} ${\rm e}_0$ &\hspace{0.2cm} $\bar{\rm e}_0$ 
&\hspace{0.2cm}  $m_A$ & \hspace{0.2cm}
$\bar{m}_A$ &\hspace{0.2cm} $\Delta_0$ &\hspace{0.2cm} $\bar{\Delta}_0$  \\
\noalign{\smallskip}
\hline\noalign{\smallskip}
$\left( 1,\frac{1}{2}\right)$  &\hspace{0.2cm} -1.45432 &\hspace{0.2cm}
 -1.45408 &\hspace{0.2cm} 0.79388 &\hspace{0.2cm} 0.79248 &\hspace{0.2cm}
1.7744 &\hspace{0.2cm} 1.76 \\
$\left( \frac{3}{2},1\right)$  &\hspace{0.2cm} -3.86321 &\hspace{0.2cm}
 -3.86192 &\hspace{0.2cm} 1.14617 &\hspace{0.2cm} 1.14427 &\hspace{0.2cm}
1.6381 &\hspace{0.2cm} 1.63 \\
$\left(\frac{3}{2},\frac{1}{2}\right)$ &\hspace{0.2cm} -1.96699
 &\hspace{0.2cm} -1.96727 &\hspace{0.2cm} 1.35666 &\hspace{0.2cm}
1.35742 &\hspace{0.2cm} 1.4217 & \hspace{0.2cm} 1.42\\
$\left( 2,\frac{1}{2}\right)$ &\hspace{0.2cm} -2.47414 &\hspace{0.2cm}
  &\hspace{0.2cm} 1.88984 &\hspace{0.2cm}  &\hspace{0.2cm} 1.2938
 &\hspace{0.2cm} 1.29\\ 
\noalign{\smallskip}\hline
\end{tabular}
\label{t1}
\end{table}

Finally, in Table~\ref{t1}  we  have collected  SWT results for
different ferrimagnetic chains. It is interesting to note
that even in the extreme quantum cases $(1,\frac{1}{2})$ and
$(\frac{3}{2},1)$, deviations from the DMRG estimates 
are less than $0.03\%$ for the energy and $0.2\%$ for the
on-site magnetization. Moreover, it is seen that the increase of
$r_s=s_1/s_2$ -- keeping $s_2=\frac{1}{2}$ fixed  --  leads to a rapid
improvement of the $1/S$ series. The above results
suggest that the  Heisenberg ferrimagnetic chains and ladders
are  examples of low-dimensional quantum spin systems where
the spin-wave approach is  an effective theoretical tool.  
\section{ Applications to  2D Heisenberg Antiferromagnets}\label{sec:4}
In this Sect.  we survey  recent applications of
the spin-wave approach to 2D Heisenberg spin systems, 
the emphasis being on the ground-state parameters 
of the square- and triangular-lattice Heisenberg antiferromagnets. 
We shall skip most of the technical details,  as the discussed 
spin-wave formalism actually does not depend on the  space dimension. 
As already mentioned, for the last decade SWT has been 
found to produce surprisingly accurate results for the 
ground-state parameters of the 
square-lattice Heisenberg antiferromagnet
even in the extreme quantum limit $S=\frac{1}{2}$. Below we collect
these results and compare them with recent QMC
numerical estimates. As to the triangular antiferromagnet,
it seems to be  a rare example of  magnetically frustrated
spin system where the spin-wave expansion is effective.
In this case,  we  also give some technical details concerning
the spin-wave expansion, as it includes some new issues resulting
from the  coplanar arrangement of classical spins.

\subsection{Square-Lattice  Antiferromagnet}\label{sec:4.1}
The square-lattice $S=\frac{1}{2}$ Heisenberg antiferromagnet --
being a simple and rather general model to describe the undoped 
copper-oxide materials -- has received  a great deal
of interest for the last decade. 
Now it is widely  accepted that the ground state
of the model is characterized by antiferromagnetic long-rage order.
Thus, the role of quantum spin fluctuations 
is restricted to reduction of the  sublattice magnetization from 
its classical value $\frac{1}{2}$ by about $39\%$.~\footnote{Compare 
with the reduction of about  $42\%$
of the classical on-site  magnetization $\frac{1}{2}$ in the 
$(1,\frac{1}{2})$ ferrimagnetic chain (see Table 1).}
In a seminal work by Chakravarty, Halperin, and 
Nelson \cite{chakravarty} -- using the renormalization-group approach
to study the quantum non-linear $\sigma$ model in $2+1$ space-time
dimension -- it has been shown that in the so-called 
renormalized classical regime $k_B T\ll \rho_s$ the  
thermodynamic properties of the 
2D quantum Heisenberg antiferromagnet  are dominated by
magnon excitations, so that the leading and next-to-leading corrections
in $k_B T / \rho_s$  are fully controlled by three physical 
parameters, i.e  the  
spin stiffness constant
$\rho_s$,\footnote{This quantity, 
measuring the response of the system to an infinitesimal 
twist of the spins around an axis perpendicular
to the direction of the broken symmetry,
should not be confused with the spin stiffness constant 
of the ferromagnetic state $\varrho_s$
connected to the Landau--Lifshitz relation (\ref{landau}).}
the  spin-wave velocity
 $c$, and the on-site  magnetization $m$, 
calculated at $T=0$ (see also \cite{hasenfratz1}). 
Moreover, it has been argued that the discussed  universal thermodynamic 
properties  appear  for arbitrary $k_B T / \rho_s$, 
provided that $0< \rho_s\ll J$ and  $k_BT\ll J$, $J$ being the 
nearest-neighbor exchange constant \cite{chubukov1}. 

The quantities $\rho_s$, and $c$   appear as input
parameters in the quantum non-linear $\sigma$ model defined
by the Lagrangian density 
\begin{equation}\label{lag1}
{\cal L} = \frac{\rho_s}{2c^2}\left( \frac{\partial {\vec n}}
{\partial t}\right)^2
-\frac{\rho_s}{2}\left[\left(\frac{\partial {\vec n}}{\partial x}\right)^2 
+\left( \frac{\partial {\vec n}}{\partial y}\right)^2\right]\, ,
\end{equation}
where the vector staggered field $\vec{n}=\vec{n}(t,x,y)$ 
satisfies the non-linear constraint ${\vec n}^2 =1$.
This model may  be introduced using arguments 
based on  general grounds:
As long as the continuous $O(3)$ symmetry is spontaneously broken,
the symmetry of the problem requires that the interaction of the
Goldstone modes, i.e. spin waves,  of the system in the 
long-wavelength limit be described 
by this model regardless of the details of the macroscopic Hamiltonian
and the value of the spin. 
For the square-lattice  antiferromagnet,  close to 
$\vec{k}=(0,0)$ and $(\pi , \pi)$ the magnon spectrum 
takes the relativistic forms 
$E_{\vec{k}}=c|\vec{k}|$ and $|\vec{\pi}-\vec{k}|$, $c$ being the
 spin-wave velocity.
 If we expand $\vec n$ as $(1,\epsilon_1 , \epsilon_2)$, where the 
$\epsilon_i$ are small compared to  unity, then the equations 
of motion following 
from (\ref{lag1}) show that there are two modes both of which have the 
dispersion $E_{\vec{k}} = c |k|$,  as expected. 
If we expand the Lagrangian to higher orders in $\epsilon_i$, we
find that there are interactions between the spin waves whose strength is
proportional to $c/\rho_s$, which is of order $1/S$.  
We thus see that all the parameters appearing in 
(\ref{lag1}) can be determined by SWT.  Compared to the standard
$1/S$ expansion, the hydrodynamic approach is more generic in
two points, i.e. (i) it is applicable to
magnetically disordered phases, and (ii) it may  lead 
to non-perturbative results which are 
beyond the reach of SWT (see, e.g. \cite{affleck3,fradkin,sachdev1}).

Ground-state parameters  of 
the $S=\frac{1}{2}$ square-lattice Heisenberg 
antiferromagnet have been studied 
in great detail  using a variety  of techniques, including SWT, 
QMC, and series expansions \cite{manousakis}. An early  QMC study by 
Reger and Young \cite{reger} indicated that the SWT gives a good
quantitative description of the ground state. Series expansions 
around the Ising limit performed by Singh \cite{singh1,singh2} found the
results $\rho_s\approx 0.18 J$ and $c\approx 1.7J$, both in
good agreement with the first-order SWT \cite{oguchi}.
Later on, higher-order calculations demonstrated that the
second-order corrections in $1/S$ to   the parameters 
$\rho_s$, $c$ and $m$  are small -- even
in the extreme quantum limit $S=\frac{1}{2}$ -- and improve 
the SWT results. For instance -- using both the Dyson--Maleev and
Holstein--Primakoff formalisms up to second order in $1/S$ -- 
Hamer et al. calculated the ground-state energy $E_0/N$ and  
the sublattice magnetization $m$ \cite{hamer1}. Both formalisms were shown
to give identical results closely approximating previous 
series estimates \cite{weihong1}. Different scientific groups have
presented consistent second-order SWT results for the spin-wave 
velocity $c$ \cite{canali2,igarashi1,weihong2},
the uniform transverse  susceptibility 
$\chi_{\perp}$\cite{igarashi1,hamer2}
 and the spin stiffness constant $\rho_s$\footnote{The reported
third-order SWT result for this parameter
 is 0.1750(1)\cite{hamer2}.}
 \cite{igarashi1,hamer2}. 
In Table~\ref{t2} we
have collected  some of these results, demonstrating an excellent 
agreement with  recent  high-precision numerical
estimates \cite{sandvik} obtained by using the stochastic 
series expansion QMC method  for $L\times L$ lattices with 
$L$ up to $16$.        
\begin{table}
\centering
\caption{
Second-order SWT results for the ground-state energy
per site  ${\rm e}_0=E_0/N$ \cite{hamer1},  the on-site  
magnetization $m$ \cite{hamer1,igarashi1},
 the spin-wave velocity $c$ \cite{igarashi1,weihong2}, the  uniform 
transverse susceptibility 
$\chi_{\perp}$  \cite{igarashi1,hamer2}, 
and the spin stiffness 
constant
$\rho_s$ \cite{igarashi1,hamer2} of the
$S=\frac{1}{2}$ square-lattice Heisenberg antiferromagnet. The 
SWT results are compared to recent stochastic 
series expansion QMC estimates  for $L\times L$ lattices 
with $L$ up to $16$ \cite{sandvik}. The series risults for ${\rm e}_0$,
$m$ and $\chi_{\perp}$ are taken from \cite{zheng}, and those
for $\rho_s$ and $c$ -- from  \cite{hamer2}.
The figures in parentheses 
show the errors in the last significant figure. $\hbar =a_0=J=1$.
}
\label{t2}
\begin{tabular}{llll}
\hline\noalign{\smallskip}
  Quantity   & \hspace{0.5cm}  SWT
  &\hspace{0.5cm}   QMC & \hspace{0.5cm} Series  \\
\noalign{\smallskip}\hline\noalign{\smallskip}
$-{\rm e}_0$ &\hspace{0.5cm}   0.669494(4)   
&\hspace{0.5cm}  0.669437(5) & \hspace{0.5cm} 0.6693(1) \\
$m$     & \hspace{0.5cm}  0.3069(1)      
&\hspace{0.5cm}   0.3070(3)&\hspace{0.5cm} 0.307(1) \\
$c$     &\hspace{0.5cm}   1.66802(3)     &  \hspace{0.5cm} 1.673(7)
&\hspace{0.5cm} 1.655(12)  \\
$\chi_{\perp}$  & \hspace{0.5cm}  0.06291(1)     
& \hspace{0.5cm}  0.0625(9)&\hspace{0.5cm} 0.0659(10)  \\
$\rho_s$ &\hspace{0.5cm}   0.180978
     & \hspace{0.5cm}  0.175(2)
&\hspace{0.5cm} 0.182(5) \\
\noalign{\smallskip}\hline
\end{tabular}
\end{table} 

The  accuracy of SWT  
may be understood  in terms  of the spin-wave interaction $V$. 
Indeed, let us consider  the $1/S$ series for
$m$ \cite{hamer1}
\begin{equation}\label{ser:m}  
m=S-0.1966019+\frac{0.003464}{(2S)^2}+{\cal O}\left(\frac{1}{S^3}\right)\, .  
\end{equation}
For $S=\frac{1}{2}$, the related series in powers of $\lambda$ simply reads
$m=0.3033981\lambda^0+0.003464\lambda^2+{\cal O}(\lambda^3)$, so that
the spin-wave interaction $V$ introduces numerically small corrections
to the leading approximation. The same conclusion is valid for
the other parameters. 
\subsection{Triangular-Lattice  Antiferromagnet}\label{sec:4.2}
The   Heisenberg antiferromagnet on a  
triangular lattice
with nearest-neighbor exchange interactions
is a typical example of strongly frustrated spin model.\footnote{For
 a recent review on frustrated quantum magnets, see \cite{lhuillier}.}

After a long period of intensive studies --
see, e.g. \cite{bernu1} and references therein --
it is now widely  accepted that the classical coplanar
ground state survives quantum fluctuations. This state
may be represented  by the ansatz
\begin{equation}\label{str} 
\frac{\vec{s}_{\vec{r}}}{S}=\hat{\vec{z}}\cos (\vec{q}_M\cdot\vec{r})
+\hat{\vec{x}}\sin (\vec{q}_M\cdot\vec{r})\, ,
\end{equation} 
where  $\vec{q}_M=(\frac{4\pi}{3},0)$ is the  wave vector 
of the magnetic pattern,  $\hat{\vec{x}}\perp \hat{\vec{z}}$ 
are unit coordinate vectors in the spin space,  and 
$\vec{r}$ runs on the lattice  sites.
As usual, the lattice spacing $a_0$ is set to unity.
The classical spins  lay in the  $(x,z)$ plane, and
point in three different directions  so that
the angle $\frac{2\pi}{3}$ is settled  between any pair of spins  
in the  elementary triangle ($\vec{s}_a,\vec{s}_b,\vec{s}_c $).  

In performing the $1/S$ expansion about non-collinear reference states
such as (\ref{str}), one faces some novelties which will be
discussed in the remainder of this Sect. One of them  concerns
the number of  boson fields needed to keep track of the whole 
magnon spectrum. This is an important practical issue,
as higher-order spin-wave expansions  involving more than two
boson fields are, as a rule,  technically intractable.
In the general case, this number should be
equal  to the number of spins in the magnetic elementary cell,
so that for the magnetic structure (\ref{str}) we would need  
three  boson fields. However, in several  special cases
we can transform  the non-collinear magnetic structures 
into a  ferromagnetic configuration  by applying a uniform 
twist on the  coordinate frame.
These special systems  have the property that 
their  magnon spectrum has no gaps at the boundaries of
the reduced magnetic Brillouin zone connected to the magnetic pattern.
The triangular-lattice antiferromagnet fulfills this rule, so that we may
describe the system by a single boson field, as in the ferromagnetic
case. In the remainder of this Sect. we shall follow this 
approach \cite{chubukov2}. 

To that end,  let us  rotate 
the spin coordinate  frame  about the $y$  axis by the angle 
$\theta_{\vec{r} \vec{r}^{'}}  = 
\vec{q}_M\cdot (\vec{r}- \vec{r}^{'})$ 
for any pair of neighboring spins 
$(\vec{s}_{\vec r},\vec{s}_{\vec{r}^{'}})$, 
in accord  to the reference state  (\ref{str}). In the  local 
reference frame, the Heisenberg Hamiltonian acquires the form 
\begin{equation}\label{htr}
{\cal H} =\! \sum_{(\vec{r},\vec{r}^{'})}\! \left[ 
\cos \theta_{\vec{r} \vec{r}^{'}}
\left( s_{\vec{r}}^x s_{\vec{r}^{'}}^x\! +\! s_{\vec{r}}^z s_{\vec{r}^{'}}^z
\right)
\!+\! \sin \theta_{\vec{r} \vec{r}^{'}}
\left( s_{\vec{r}}^z s_{\vec{r}^{'}}^x \! -\! s_{\vec{r}}^x s_{\vec{r}^{'}}^z
\right) \! +\! s_{\vec{r}}^y s_{\vec{r}^{'}}^y\right] ,
\end{equation}
where the sum runs  over  all pairs  of nearest-neighbor 
sites of the triangular lattice. 

Next, using the Holstein--Primakoff transformation (\ref{hpr})\footnote{
The choice of the transformation is a matter of convenience, as the
final results -- at least to second order in $1/S$ -- are independent
of the boson representation.} and 
the procedures described in Sect. 2, we find the following boson
representation for (\ref{htr})
\begin{equation}\label{hbtr}
{\cal H}_B=-\frac{3}{2}S^2N+3S\sum_k\left[
A_{\vec{k}}a_{\vec{k}}^{\dag}a_{\vec{k}}+\frac{B_{\vec{k}}}{2}
\left( a_{\vec{k}}^{\dag}a_{-\vec{k}}^{\dag}+
a_{\vec{k}}a_{-\vec{k}}\right)\right] +V\, ,
\end{equation}
$A_{\vec{k}}=1+\nu_{\vec{k}}/2$, 
$B_{\vec{k}}=-3\nu_{\vec{k}}/2$, and
$\nu_{\vec{k}}=\frac{1}{3}[\cos k_x+2\cos (k_x/2)\cos (\sqrt{3}k_y/2)]$.
Here and in the remainder of this Sect.,
$\vec k$ takes $N$ values from the first Brillouin zone of 
the triangular lattice.
 
Up to quartic anharmonic terms, the expansion of the square root 
in  (\ref{htr}) produces the following spin-wave interaction 
$V=V_3+V_4$, where
\begin{eqnarray}\label{v:tr}
V_3&=&\imag  \sqrt{\frac{S}{2}}\frac{3}{2\sqrt{N}}
\sum_{1-3}(\kappa_1+\kappa_2)
(a_1^{\dag}a_2^{\dag}a_3-a_3^{\dag}a_2a_1)\, ,\\
V_4&=&-\frac{3}{16N}\sum_{1-4}\left[ \Gamma_{12;34}^{(1)}
a_1^{\dag}a_2^{\dag}a_3a_4+\Gamma_{123}^{(2)}
(a_1^{\dag}a_2^{\dag}a_3^{\dag}a_4+a_4^{\dag}a_3a_2a_1)\right]\, ,
\end{eqnarray}
$\kappa_{\vec k}=\frac{1}{3}[\sin k_x-2\sin (k_x/2)\cos (\sqrt{3}k_y/2)]$,
$\Gamma_{12;34}^{(1)}=4\nu_{1-3}+4\nu_{2-3}+\nu_1+\nu_2+\nu_3+\nu_4$,
and $\Gamma_{123}^{(2)}=-2(\nu_1+\nu_2+\nu_3)$. 
For simplicity, in the last expressions we have omitted the
Kronecker $\delta$ function, and have used the  abbreviations for the
wave vectors introduced in Sect.~\ref{sec:2.2}.

A novelty here is the triple boson interaction $V_3={\cal O}(S^{1/2})$,
which is  typical for systems exhibiting non-collinear magnetic
patterns. We shall see below that such kind of interactions complicate
the calculation of higher-order   $1/S$ corrections.
\subsubsection{Linear Spin Wave Approximation}\label{sec:4.2.1}
In a standard LSWT, we discard  $V$ 
and diagonalize the quadratic part of (\ref{hbtr}) by the
Bogoliubov transformation $a_{\vec k}=u_{\vec k}(\alpha_{\vec k}
-x_{\vec k}\alpha_{-\vec k}^{\dag})$.  The parameters 
 $u_{\vec k}$ and $x_{\vec k}$ are defined
by (\ref{uv}) and (\ref{par}), but in this case 
$\eta_{\vec k}=-3\nu_{\vec k}/(2+\nu_{\vec k})$. The diagonalization
yields the free-quasiparticle Hamiltonian  
${\cal H}_0=3S\sum_{\vec k}\omega_{\vec k} 
\alpha_{\vec k}^{\dag}\alpha_{\vec k} $, where the dispersion
function
\begin{equation}\label{sp:tr}
E_{\vec k}\equiv 3S\omega_{\vec k} =3S\sqrt{(1-\nu_{\vec k})(1+
2\nu_{\vec k})}
\end{equation}
gives the magnon energies in a LSWT approximation, to be compared 
with the magnon spectrum resulting from the approach using
three boson fields \cite{jolicoeur1}. It is easy to check that
the dispersion function (\ref{sp:tr}) exhibits three zero modes,
as it should be since the Hamiltonian symmetry $O(3)$ is 
completely broken by the magnetic pattern (\ref{str}).
Two of these modes are at the ordering wave vectors  $\vec{k}=\pm \vec{q}_M$,
whereas the third zero mode at $\vec{k}=0$ describes soft fluctuations
of the total magnetization. Expanding about the zero modes, we
find the following expressions for the spin-wave velocities \cite{dombre}
\begin{equation}\label{c:tr}
c_{0\perp}\equiv c_{\pm q_M}=\left(\frac{3}{2}\right)^{3/2}S\, ,
\hspace{0.5cm} c_{0\|}\equiv c_{k=0}=\frac{3\sqrt{3}}{2}S\, .
\end{equation}

Let us now calculate the on-site magnetization $m=\langle
s_{\vec r}^z\rangle=S-\langle a_{\vec r}^{\dag}a_{\vec r}\rangle$.
Using the Bogoliubov transformation, we find for the density
of particles $\langle a_{\vec k}^{\dag}a_{\vec k}\rangle =-1/2
+1/(2\varepsilon_{\vec k})$, so that the LSWT result for $m$ 
reads \cite{jolicoeur1}
\begin{equation}\label{m:tr} 
m=S+\frac{1}{2}-\frac{1}{2N}\sum_{\vec k}
\frac{1}{\sqrt{1-\eta_{\vec k}^2}}= S-0.2613\, .
\end{equation}

For $S=\frac{1}{2}$, the LSWT result
is  $m = 0.2387$.  Since the reported leading $1/S$ 
correction to $m$ is small and positive\footnote{We are aware of two such 
calculations reporting, however,  somewhat different corrections, i.e. 
 $0.0055/S$ \cite{miyake1} and $0.00135/S$ \cite{chubukov2}.},
there is  a clear disagreement with the  recent QMC estimate 
$m=0.20(6)$ \cite{capriotti}.
\subsubsection{Spin Wave Interactions}\label{sec:4.2.2}
Here we consider as an example the calculation of   $1/S$
corrections to the magnon spectrum (\ref{sp:tr}). 
There are two different types of corrections related to 
the spin-wave interactions $V_3$ and $V_4$ in  (\ref{v:tr}).
Turning to $V_4$, notice that we have already learned
(Sect.~\ref{sec:2.3}) that the required correction  may be 
obtained  by  expressing $V_4$ as a sum of 
 normal products of quasiparticle 
operators: the diagonal quadratic terms  give the  
 required $1/S$ correction to the spectrum. 
However, in several  cases we are not interested in 
the quasiparticle representation  of
$V_4$. Then,  it is possible  to follow another 
way  by  decoupling the quartic operator products  in $V_4$.
Actually, this  procedure takes 
into account the so-called  one-loop diagrams, and may be
performed within a formal substitution 
of the operator products, such as  $a_1^{\dag}a_2^{\dag}a_3a_4$, 
by the following sum over  all the non-zero pair boson
correlators 
\begin{equation}\label{pair}
a_1^{\dag}a_2^{\dag}a_3a_4\longmapsto 
\sum_{\rm pair}
\left[ \langle  a_1^{\dag}a_2^{\dag}\rangle a_3a_4
+a_1^{\dag}a_2^{\dag} \langle   a_3a_4 \rangle
-\langle  a_1^{\dag}a_2^{\dag}\rangle  \langle   a_3a_4 \rangle
\right]\, .
\end{equation}
As suggested by  the quadratic form in (\ref{hbtr}), 
there are two types of  boson  correlators,  i.e.  
$\langle  a_1^{\dag}a_2\rangle$
and $\langle  a_1a_2\rangle = \langle  a_1^{\dag}a_2^{\dag}\rangle$,
contributing in (\ref{pair}). 
The constant terms in (\ref{pair})
give  first-order corrections to the ground state energy, 
whereas the quadratic operator products renormalize the coefficients 
$A_{\vec k}$ and  $B_{\vec k}$  in (\ref{hbtr}). 
Thus, the interaction $V_4$ renormalizes  the bare dispersion function 
$\omega_{\vec k}$ to
\begin{equation}\label{ren:sp}
\bar{\omega}_{\vec k}=\sqrt{\bar{A}_{\vec k}^2-
\bar{B}_{\vec k}^2}\, ,
\end{equation}
where the new  coefficients $\bar{A}_{\vec k}$ and 
$\bar{B}_{\vec k}$ can be expressed in  the form\footnote{For brevity, 
here we omit  the  expressions for 
the constants $a_1$, $a_2$, $b_1$, and $b_2$ \cite{chubukov2}.} 
$$
\bar{A}_{\vec k}=A_{\vec k}\left(1+\frac{a_1}{2S}\right)
+\frac{a_2}{2S}\, ,\hspace{0.5cm}
\bar{B}_{\vec k}=B_{\vec k}\left(1+\frac{b_1}{2S}\right)
+\frac{b_2}{2S}\, .
$$

An  analysis of (\ref{ren:sp}) indicates  that the renormalized spectrum 
still preserves the zero mode at $\vec{k}=0$, but 
at the same time acquires  non-physical gaps  at 
$\vec{k}=\pm \vec{q}_M$. The reason for such kind of behavior
of the SWT  is connected with the fact that we have omitted
the $1/S$ corrections resulting from $V_3$. Indeed, 
the spin-wave interaction $V_3$ has the order ${\cal O}(S^{1/2})$, 
so that a simple power counting indicates  that  
$1/S$ corrections  to $\omega_{\vec k}$  appear in the 
second-order of the perturbation theory in $V_3$. 
We shall skip the details of this calculation, 
as it   may be performed entirely
in the framework of the method presented in Sect.~\ref{sec:2}. Namely,
one  should express $V_3$ in terms of quasiparticle operators, and 
then apply  the general  perturbation 
formula (\ref{form1}) for the interaction $V_3$, 
by using  the dressed dispersions (\ref{ren:sp}). 
As a matter of fact, as  we are interested in corrections up to $1/S$, 
we can  use the bare dispersion function (\ref{sp:tr}).  
The final  result of this calculation 
shows that the $1/S$ correction resulting from $V_3$
exactly  vanishes  the gap (produced by $V_4$),  so that
the structure of  magnon spectrum (\ref{sp:tr}) -- 
containing three zero modes -- is  preserved 
in the leading first-order approximation \cite{chubukov4}. 
Based on  the renormalized  dispersion, the following
expressions for  the spin-wave velocities (\ref{c:tr})
have been reported \cite{chubukov2}:
$$
c_{\|}=c_{0\|}\left(1-\frac{0.115}{2S}\right)\, ,\hspace{0.5cm}
c_{\perp}=c_{0\perp}\left(1+\frac{0.083}{2S}\right)\, .
$$
Notice that the $1/S$ corrections  diminish the ratio
$c_{\|}/c_{\perp}$ from the LSWT result  $1.41$ to the value $1.16$.
These expressions indicate  that the leading corrections
to  the magnon spectrum are numerically small even
in the case $S=\frac{1}{2}$. Good convergence  has  been  
found also for  the $1/S$ series of  the magnetic susceptibilities
$\chi_{\perp}$ and $\chi_{\|}$ \cite{chubukov5,trumper1} which
appear as  parameters of the magnetic susceptibility 
tensor \cite{andreev}
$$
\chi_{\alpha\beta}=\chi_{\perp}\delta_{\alpha\beta}+
(\chi_{\|}-\chi_{\perp})y_{\alpha})y_{\beta}\, .
$$
Here  $\hat{\vec y}$ is a unit vector directed perpendicular 
to the basal ($x,z$) plane of the planar magnetic structure.  

Summarizing, the available SWT results point towards a 
good convergence of the perturbative spin-wave series in 
the triangular-lattice Heisenberg antiferromagnet.   
This is  remarkable, as  the spin-wave expansion
might have been expected to fail for
strongly  frustrated magnetic systems. 
\section{Modified  Spin Wave Theories}\label{sec:5}
Here  we consider some modifications of the
standard spin-wave theory allowing for an analysis of magnetically 
disordered phases. These may appear either as a result of quantum
fluctuations -- a  classical example being the spin-$S$
Heisenberg antiferromagnetic chain discussed
in Sect.~\ref{sec:3.1} -- or due to thermal fluctuations,
as in $1D$ and $2D$ Heisenberg  magnets with short-range
isotropic interactions \cite{mermin}. For the antiferromagnetic chain, 
we have indicated that the failure of
SWT arises already in the linear spin-wave approximation  
as a  divergency  in the boson-occupation numbers 
$n_i=\langle a_i^{\dag}a_i\rangle =\infty$ implying 
$\langle s_i^z\rangle =-\infty$ . Infinite number of spin waves also 
appears at $T>0$,   when the $T=0$  magnetic phases 
of low-dimensional  Heisenberg systems  do not survive thermal fluctuations.
Actually, the occupation numbers $n_i$ should not exceed $2S$ --  as
dictated by the spin algebra --  and the magnetization should be zero,
as required by the symmetry of the phases.
In the remainder of this Sect.  we discuss modifications of the SWT
based on  \emph{ad hoc} constraints imposing fixed number of
bosons.

The first generalized spin-wave theory  of this 
kind has been  formulated
by Takahashi to study  the    low-$T$ thermodynamics of $1D$ and 
$2D$ Heisenberg ferromagnets \cite{takahashi1,takahashi2}. 
Takahashi's idea was to supplement the standard SWT of 
Heisenberg ferromagnets
with the constraint imposing zero ferromagnetic moment at $T>0$:
\begin{equation}\label{constr:eq}
M=\sum_{n=1}^N\langle s_n^z\rangle=SN-\sum_{\vec k}
\langle a_{\vec k}^{\dag} a_{\vec k}\rangle =0\, .
\end{equation}
Depending on the context, in the remainder of this Sect.
 $\langle A \rangle$ means the expectation 
value of the  operator $A$ at $T=0$ or  $T>0$. 
Quite surprisingly, it was  found  an excellent agreement 
with the  Bethe-ansatz low-temperature expansions 
of  the  free energy  and magnetic susceptibility  for the 
$S=\frac{1}{2}$ Heisenberg ferromagnetic chain. 
Similar extensions of SWT have been suggested for Heisenberg
antiferromagnets both at $T=0$ \cite{hirsch,zhong} 
and at $T>0$ \cite{takahashi3,arovas},  by  using
the  same constraint equation (\ref{constr:eq}) but
for the  sublattice magnetization.  Below we discuss some
applications of the modified SWT to low-dimensional
Heisenberg antiferromagnets both at $T=0$ and at finite
temperatures.  
\subsection{Square-Lattice Antiferromagnet at Finite $T$}\label{sec:5.1}
Using the Dyson--Maleev transformations (\ref{dma}) and   (\ref{dmb}),
 the boson  Hamiltonian ${\cal H}_B^{'}$ of the
 square-lattice antiferromagnet  reads
\begin{equation}\label{h:sqare}
{\cal H}_B^{'}\! =\! -\frac{N}{2}JzS^2\! +\sum_{\vec k}
\left[ A_{\vec k}(a_{\vec k}^{\dag} a_{\vec k}\! 
+\! b_{\vec k}^{\dag} b_{\vec k})
\! +\! B_{\vec k}(a_{\vec k}^{\dag}b_{\vec k}^{\dag}\!+\!
a_{\vec k} b_{\vec k})\right]\!+\!V_{DM}^{'}\, ,
\end{equation}
whereas the constraint equation for the total sublattice
magnetization takes the form
\begin{equation}\label{c:eq}
\sum_{\vec k}\langle  a_{\vec k}^{\dag} a_{\vec k}+
b_{\vec k}^{\dag} b_{\vec k}\rangle =SN\, .
\end{equation}
The wave vector $\vec k$ runs in the small (magnetic) 
Brillouin zone $|k_x\pm k_y|\leq \pi$ containing $N/2$ points.
$A_{\vec k}=JSz$, $B_{\vec k}=JSz\gamma_{\vec k}$,
$\gamma_{\vec k}=\frac{1}{2}(\cos k_x+\cos k_y)$, and
$z=4$ is the lattice coordination number.

In essence, the constraint equation (\ref{c:eq}) introduces an
effective cut-off for unphysical states \cite{dotsenko}. 
To see this, let us
consider the $S=\frac{1}{2}$ system. According to  (\ref{c:eq}),
the average number of, say, the  $\alpha$  magnons is $N/4$,
whereas the total  number of one-magnon states in the magnetic 
Brillouin zone is $N/2$. Thus, after introducing the constraint  
(\ref{c:eq}),
the effective number of allowed states in the boson Hilbert
space is
$$
\left[\frac{(N/2)!}{(N/4)!(N/4)!}\right]^2\sim \frac{4}{\pi}
\frac{2^N}{N}\, , 
$$
so that with logarithmic accuracy the correct dimension $2^N$ is
restored. 

To implement the constraint equation in the theory,
we  introduce, as usual,  a chemical potential 
$\mu$ for  the boson fields, 
i.e. instead of 
${\cal H}_B^{'}$   we consider the Hamiltonian
\begin{equation}\label{h:mu}
  {\cal H}_B= {\cal H}_B^{'}-\mu
\sum_{\vec k} (a_{\vec k}^{\dag} a_{\vec k}+
b_{\vec k}^{\dag} b_{\vec k})\, ,
\end{equation}
where $\mu$ is fixed by  the constraint
equation (\ref{c:eq}).  Notice that the introduction 
of a chemical potential simply renormalizes the coefficient 
$A_{\vec k}\rightarrow A_{\vec k}-\mu$ so that we can apply  the 
formalism from Sect.~\ref{sec:2} without any changes.

 Using the Bogoliubov  transformation (\ref{bt}) with
the parameter $\eta_{\vec k}=JzS\gamma_{\vec k}/(JzS-\mu)$, 
one finds 
the following quasiparticle representation of ${\cal H}_B$
(see, e.g. \cite{canali1})
\begin{equation}
{\cal H}_B =E_0+{\cal H}_D+V_{DM}\, , 
\end{equation}
where $E_0$ is the ground-state energy calculated up to
first-order of the perturbation theory in $1/S$:
\begin{equation}\label{e0:sq}
E_0=-\frac{N}{2}zJS^2\left(1+\frac{r}{2S}\right)^2\, ,
\hspace{0.5cm} r=1-\frac{2}{N}\sum_{\vec k}\sqrt{1-\eta_{\vec k}^2}\, .
\end{equation}
As we know  from Sect.~\ref{sec:2.3}, the free-quasiparticle
Hamiltonian
\begin{equation}
{\cal H}_D= \sum_{\vec k}E_{\vec k}(\alpha_{\vec k}^{\dag}\alpha_{\vec k}
+\beta_{\vec k}^{\dag}\beta_{\vec k})
\end{equation}
includes  the diagonal quadratic terms resulting from $V_{DM}^{'}$,
so that the magnon energies $E_{\vec k}$ are  calculated up to
first-order corrections in $1/S$:
\begin{equation}\label{sp:sqare}
E_{\vec k}=JzS\left(1+\frac{r}{2S}\right) \sqrt{1-\eta_{\vec k}^2}\, .
\end{equation}
Here the factor $r/2S$ is  Oguchi's correction to the
magnon spectrum \cite{oguchi}.

We want to  treat the spin-wave interaction 
up to  first order in  the $1/S$ perturbation theory, 
so that the  Dyson--Maleev interaction $V_{DM}$ 
will  be discarder. It is important 
to notice that here  the  off-diagonal quadratic
interaction $V_2$ does not appear, as dictated
by the sublattice interchange symmetry. This means 
that the lowest-order corrections to the sublattice 
magnetization $m$ have the order ${\cal O}(S^{-2})$, see 
the series (\ref{ser:m}), 
so that  the constraint equation (\ref{c:eq}) calculated
in a LSWT approximation can be
safety used at this level.

Turning to the magnon spectrum (\ref{sp:sqare}), we see that the
chemical potential introduces a spectral gap $\Delta$ so that close
to the zone center the excitation spectrum acquires the relativistic
form
\begin{equation}\label{ekk}
E_{\vec k}=\sqrt{\Delta^2+c^2k^2}\, , \hspace{0.5cm}
c=\frac{JzS}{\sqrt{2}}\left(1+\frac{r}{2S}\right)\, ,
\end{equation}
where $\Delta =2c(-\mu /JzS)^{1/2}$ and $c$ is the spin-wave
velocity calculated up to first order in $1/S$. 
Using the standard expression for free bosons $n_{\vec k}=
\langle \alpha_{\vec k}^{\dag} \alpha_{\vec k}\rangle = 
\langle \beta_{\vec k}^{\dag} \beta_{\vec k}\rangle =
[\exp (-E_{\vec k}/k_BT)-1]^{-1}$, the constraint equation (\ref{c:eq})
takes the form
\begin{equation}\label{gap:eq} 
S+\frac{1}{2}=\frac{1}{N}\sum_{\vec k}\frac{1}{\sqrt{1-\eta_{\vec k}^2}}
\coth \frac{E_{\vec k}}{k_BT}\, .
 \end{equation}  
At low $T$,  the main contributions in the last sum 
come from small wave vectors  so that,  using  (\ref{ekk}), 
the gap equation  (\ref{gap:eq}) yields
\begin{equation}\label{gap}
\Delta=\frac{c}{\xi}=2T{\rm arcsinh} \left[ \frac{1}{2}\exp\left(
-\frac{2\pi\rho_s}{k_BT}\right)\right]\, . 
\end{equation}    
Here $\rho_s$ is the  $T=0$ spin stiffness constant
calculated up to first order in $1/S$, 
and $\xi$ is the spin correlation length.
This result exactly reproduces the saddle-point equation in the $1/N$
expansion of the $O(N)$ nonlinear $\sigma$ model in $2+1$ space-time
dimensions (see, e.g. \cite{tsvelik}). It is well known that (\ref{gap})
describes three different regimes, i.e. (i) the  renormalized classical,
(ii) the quantum critical, and (iii) the quantum disordered regimes 
\cite{sachdev1}.

As an example, we  consider the renormalized classical
regime  defined by the 
condition $k_B T\ll\rho_s$. In this case, the last equation
yields the following result for  the correlation length
\begin{equation}\label{c:length}
\xi\sim \frac{c}{T}\exp \left(\frac{2\pi\rho_s}{k_BT}\right)\, .
\end{equation}
This  coincides with the one-loop approximation of the
$2+1$ nonlinear $\sigma$ model \cite{chakravarty}.
As is well known, at a  two-loop level  the
$T$ dependence in the pre-exponential factor disappears, whereas the
exponent argument does not  change.

Finally, let us calculate the leading temperature correction to
the internal energy $U=\langle {\cal H}_B\rangle$.
The expression for $U$ reads
\begin{equation}\label{int:en}
U=E_0+\sum_{\vec k}E_{\vec k}
\left(\coth \frac{E_{\vec k}}{k_BT}-1\right)\, .
\end{equation}
Using (\ref{ekk}), 
after some algebra one  finds the following   
result:
\begin{equation}\label{int:en2}
U=E_0
+\frac{2\zeta (3)N}{\pi c^2}T^3\, .
\end{equation}
Here $\zeta (x)$ is the Riemann zeta function. 
The above temperature correction  describes the  contribution 
from  two  zero  modes, i.e.  $\vec{k}=(0,0)$ and $\vec{k}=(\pi,\pi)$,  
and   reproduces the expected universal behavior known 
from  the $2+1$ nonlinear $\sigma$ model and 
the chiral perturbation theory \cite{hasenfratz1,hasenfratz2}.

\subsection{Applications to Finite-Size Systems}\label{sec:5.2}
The modified  SWT can also be applied to finite-size
systems \cite{hirsch,zhong}. This opens an  opportunity
directly to compare SWT results  with finite-size numerical data.   
As is known, the standard SWT is not applicable to finite
systems due to divergences related  to the
Goldstone zero modes. Actually, the divergency comes
from the  Bogoliubov transformation (\ref{bt}) which is 
not defined for these modes.

Turning to the example from  Sect.~\ref{sec:5.1}, notice that
in the infinite system the chemical potential $\mu$ 
goes to zero as $T\rightarrow 0$.
At $T=0$   the constraint equation takes  the
form  
\begin{equation}\label{gap0:eq} 
S+\frac{1}{2}-\frac{2}{N\sqrt{1-\eta_{0}^2}}
-\frac{1}{N}\sum_{\vec{k}\neq 0}
\frac{1}{\sqrt{1-\eta_{\vec k}^2}}=0\, .
\end{equation}  
Here we have selected the contribution from the  two zero modes
at $\vec{k}=(0,0)$ having $S^z=\pm 1$.  
 
According to (\ref{gap0:eq}), on a finite lattice
the parameter  $\eta_{0}=JzS/(JzS-\mu)$
is less than unity, so that the  
divergences associated
with the zero modes disappear.
The  constraint (\ref{gap0:eq})
 takes into account  the fact that in finite systems
there are no  spontaneously broken continuous symmetries.      

To find the staggered magnetization $m$ appearing 
in the thermodynamic limit of the $2D$ system, we calculate the 
antiferromagnetic structure factor $S(\pi,\pi)$ for large $N$: 
\begin{equation}\label{str:fact} 
m^2(N)=\frac{2}{N} S(\pi,\pi)=\frac{4}{(1-\eta_0^2)N^2}
+\frac{1}{N^2}\sum_{\vec{k}\neq 0}\frac{1+\eta_{\vec k}^2}
{1-\eta_{\vec k}^2}\, ,
\end{equation} 
where we have again selected the contribution from
the zero modes.

In the large-$N$ limit, the last sum transforms into an integral 
which is $\propto \ln N$, so that the main contribution comes
from the first term in (\ref{str:fact}). Thus, we find the relation 
\begin{equation}\label{m:sq} 
m^2=\lim_{N \to\infty} \frac{4}{(1-\eta_0^2)N^2}\, .
\end{equation}

Equation  (\ref{gap0:eq}) induces a gap in the 
magnon spectrum which is  defined  by 
$\Delta =c\sqrt{2(1-\eta_0^2)}$.
Using (\ref{m:sq}) and the notations from Sect.~\ref{sec:5.1},
we find the following result  for the magnon excitation gap
in the large-$N$ limit
\begin{equation}\label{sec5:gap}
\Delta =\frac{c^2}{\rho_sL^2}\, .
\end{equation}
$L=N^{1/2}$ is the linear size in a square geometry.
The last expression  reproduces  the result for $\Delta$ 
obtained by other methods  
\cite{neuberger,fisher,hasenfratz1}.

Finally, let us   return to the 
Heisenberg antiferromagnetic chain 
discussed in Sec.~\ref{sec:3.1},  this time  using the 
modified SWT \cite{arovas}. We have  seen that in $1D$
the expression for the staggered magnetization
(\ref{measure1}) contained an infrared divergency
indicating that the magnetic order is destabilized
by quantum fluctuations. Using the concept of the modified theory,
we can resolve the problem by replacing (\ref{measure1}) 
with  the constraint equation
\begin{equation}\label{gap:1d}
S+\frac{1}{2}=\frac{1}{N}\sum_{k}\frac{1}{\sqrt{1-\eta_0^2
\cos^2 k}}=\frac{K(\eta_0)}{\pi}\, ,
\end{equation}
where   $K(\eta_0)$ is the complete elliptic integral of the
first kind.

Since $K(\eta_0)\geq \pi/2$, the gap equation  (\ref{gap:1d})
has a solution for arbitrary $S$. However, the constraint 
introduces  an excitation gap, so that the discussed theory
makes sense  only for integer $S$. To find the gap,
we may  use for small $(1-\eta_0^2)^{1/2}$ the asymptotic result 
$K(\eta_0)=\ln 4(1-\eta_0^2)^{-1/2}$, so that the excitation gap
reads
\begin{equation}\label{gap:1dd}
\Delta\sim c\, \exp\, (-\pi S)\, .
\end{equation}
Here  $c$ is the spin-wave velocity of the antiferromagnetic
chain (\ref{af:sp}).
The obtained gap has the asymptotic form $\Delta\sim S\exp (-\pi S)$,
to be compared with Haldane's result   
$\Delta\sim S^2\exp (-\pi S)$ obtained from the $\sigma$-model
mapping \cite{haldane1,haldane2}. It is remarkable that
the  simple  modified SWT is capble to  reproduce 
the asymptotic expression for the Haldane gap.
\section{Concluding Remarks}\label{sec:6}
We have surveyed  the spin-wave technique
and its typical applications to Heisenberg magnetic systems 
in restricted geometries. In most of the  cases the  SWT results 
were  compared with the available numerical estimates.
As a result, the systematic large-$S$ technique has been found to 
give  very accurate description of the zero-temperature parameters
and magnon excitation spectra of a number of low-dimensional 
quantum spin models,  such as the  Heisenberg 
antiferromagnet on square and triangular lattices  and various  
quasi-one-dimensional  mixed-spin Heisenberg systems exhibiting
ferrimagnetic ground states. Presented analysis of the  asymptotic  
series up to second order  in the parameter $1/S$ implies that  
in these systems the spin-wave interaction introduces
numerically small corrections to the principal approximation,
even in the extreme quantum limit $S=\frac{1}{2}$.
Thus,  indicated  effectiveness   of the 
spin-wave technique --  as applied to  magnetic  systems 
with  small spin quantum numbers and in
restricted geometries --  may be attributed  to the 
observed  weakness  of spin-wave interactions.
  
The authors thank J. Richter  and U. Schollw\"ock  for 
their collaborations in this field, and  S. Sachdev,  A.W. Sandvik, and
Z. Weihong for the  permission to use their results.
This work was supported by the Deutsche 
Forschungsgemeinschaft.  
\end{document}